\documentclass{article}

\usepackage{arxiv}

\usepackage{graphicx}
\usepackage[caption=false]{subfig}

\usepackage[utf8]{inputenc} 
\usepackage[T1]{fontenc}    
\usepackage{hyperref}       
\usepackage{url}            
\usepackage{booktabs}       
\usepackage{amsfonts}       
\usepackage{nicefrac}       
\usepackage{microtype}      
\usepackage{epstopdf, epsfig}
\usepackage{pgfplotstable}
\usepackage{booktabs}
\usepackage{filecontents}
\usepackage{longtable}
\usepackage{array}
\usepackage{amsmath} 
\usepackage{bm}
\usepackage{amsmath}%
\usepackage{MnSymbol}%
\usepackage{wasysym}%
\usepackage{setspace}
\usepackage{lineno}

\title{An Open Source, Versatile, Affordable Waves in Ice Instrument for Scientific Measurements in the Polar Regions}
\author{
  Jean Rabault\\
  Department of Mathematics\\
  University of Oslo\\
  \texttt{jean.rblt@gmail.com} \\
   \And
  Graig Sutherland\\
  Environment and Climate Change Canada, Dorval, Canada\\
  \texttt{graigory.sutherland@canada.ca} \\
  \And
  Olav Gundersen\\
  Department of Mathematics\\
  University of Oslo\\
  \texttt{olav.gundersen@mn.uio.no} \\
   \And
  Atle Jensen\\
  Department of Mathematics\\
  University of Oslo\\
  \texttt{atlej@math.uio.no} \\
   \And
  Aleksey Marchenko\\
  The University Center in Svalbard\\
  \texttt{Aleksey.Marchenko@unis.no} \\
   \And
  {\O}yvind Breivik\\
  Norwegian Meteorological Institute \& University of Bergen,
  Bergen\\
  \texttt{oyvindb@met.no} \\
}

\begin{document}
\maketitle

\begin{abstract}
  Sea ice is a major feature of the polar environments. Recent changes in
  the climate and extent of the sea ice, together with increased economic
  activity and research interest in these regions, are driving factors for
  new measurements of sea ice dynamics. Waves in ice are important as they
  participate in the coupling between the open ocean and the ice-covered
  regions. Measurements are challenging to perform due to remoteness and
  harsh environmental conditions. While progress has been made in 
  observing wave propagation in sea ice using remote methods, these are still
  relatively new measurements and would benefit from more in situ data for 
  validation. In this article, we present an open source
  instrument that was developed for performing such measurements. The
  versatile design includes an ultra-low power unit, a microcontroller-based
  logger, a small microcomputer for on-board data processing, and an Iridium
  modem for satellite communications. Virtually any sensor can be used with
  this design. In the present case, we use an Inertial Motion Unit to record
  wave motion. High quality results were obtained, which opens new
  possibilities for in situ measurements in the polar regions. Our instrument
  can be easily customized to fit many in situ measurement tasks, and we
  hope that our work will provide a framework for future developments of a
  variety of such open source instruments.
\end{abstract}


\section{Measurements of waves in ice}

The interaction between surface waves and sea ice involves
many complex physical phenomena such as viscous damping
\citep{WeberArticle, rabault_sutherland_gundersen_jensen_2017}, wave
diffraction \citep{SquireOOWASI}, and nonlinear effects in the ice
\citep{AnalysisPolarsten}. Therefore, it is complex and still an area of
ongoing research \citep{rabault2018investigation, squire2018fresh}. Better
understanding and modeling of wave propagation in sea ice can allow for
the improvement of ocean models to be used for climate, weather and sea
state predictions \citep{KaiReport}, the estimation of ice thickness
\citep{Wadhams200998}, and the analysis of pollution dispersion in the
Arctic environment \citep{Pfirman1995129,Rigor199789}. More generally, all
these aspects must be better understood to allow safe, environment-friendly
operations in the Arctic. Therefore, there is considerable interest in
measuring sea ice dynamics.

While the use of Synthetic Aperture Radar (SAR) images from satellites to
obtain spectral wave information has made significant progress lately, they are
still not standard measurements. Indeed, they require the satellite to be in a
particular sampling mode (wide swath), as well as accurate prediction of the
azimuth cutoff wavelength. This is vital for accurate determination of the
wave energy, and is made complicated by the presence of sea ice~\citep{ardhuin2017measuring, stopa2018wave}. Therefore,
in situ measurements are still a key method for measurements of waves in ice.
In addition, as remote sensing and satellite-based measurements are still
relatively new, they still benefit greatly from validation with in situ
observations. Such in situ measurements are usually
performed using tiltmeters, accelerometers, and more recently further
refinements around those devices in the form of Inertial Motion Units (IMUs),
which combine several physical measurements (acceleration, angular rates, and
magnetic field) together with signal processing capabilities in a single device
\citep{POL:5401540, POL:5398656, liu1991wave, doble_mercer_meldrum_peppe_2006,
Doble2013166, doble2017robust, MARCHENKO2019101861}. Unfortunately, the harsh environment sets demanding requirements
for scientific observations in the Arctic. Several commercial solutions are
available (companies selling instruments operating in the arctic include, e.g.,
Sea Bird Scientific Co., Campbell Scientific Co., Aanderaa Data Instruments
A.S.), but these usually have a high cost and reduced flexibility, being
closed-source black boxes. This is especially problematic as there is a broad
consensus in the community that more observations are needed to further develop
waves in ice models. In particular, modeling and parametrization of wave
attenuation by sea ice remains a topic of active research \citep{Wang201090,
JGRC:JGRC11467, Zhao201571, SUTHERLAND2019111, rabault_sutherland_jensen_christensen_marchenko_2019, MARCHENKO2019101861}. In addition, the strong spatial inhomogeneity of
the ice has proven to be important for the evaluation of wave damping and ice
melting, and studying the effect of such spatial variations in more detail will
require many instruments to be deployed simultaneously on large-scale
field-measurements campaigns, which in turns strengthens the need for
affordable, customizable instruments  \citep{horvat2016interaction,
horvat2015prognostic, roach2018emergent, hwang2017winter}.  Fortunately,
off-the shelf sensors and open source electronics are now sufficiently
documented and easy to use as to become a credible alternative to traditional
solutions. Therefore, they may offer a solution for gathering the large amounts
of data that the community needs, at a reduced cost.

In a recent work \citep{rabault_sutherland_gundersen_jensen_2017}, a simple
open-source instrument for the measurement and logging of waves in ice was
developed. This design was further improved and iterated, by adding on-board
processing and satellite communication capabilities. As a consequence, the
new instrument has become a powerful solution for performing in situ
measurements of complex physical processes such as waves in ice. In this
article, we provide a detailed technical description of this new instrument,
a brief overview of the data collected which confirms proper functionality,
a cross-validation against both data collected using another kind of
instrument and a wave model, and we discuss implications for the communities
performing in situ measurements in the polar regions. We believe that releasing
our design as open source material may help create a scientific community
sharing the design of their instruments, therefore making the study of the
Arctic much more cost effective and enabling far more data to be collected.

\section{Technical solution used}

In this section, we present a technical description of the design and
performance of the waves in ice instrument. All the code and the files for the
Printed Circuit Board (PCB) are released as Open Source material (see Appendix
A), so that our design is fully reproducible. Moreover, the design is highly
modular, and therefore it would be easy to add sensors to the logger and to
measure and transmit additional information, such as wind speed, temperature,
humidity, irradiance, or any other relevant physical quantity that could
be required. We choose to design our instrument around affordable, easily
available and well documented hardware such as Arduinos and Raspberry Pis,
and we prefer solutions that are slightly non-optimal but easy to build,
rather than more involved designs which would require sophisticated assembly.

\subsection{Hardware}

The technical solution used in the present work is a further development of
what was presented in \cite{rabault_sutherland_gundersen_jensen_2017}. At
the core of the instrument lies a GPS and a high-accuracy, thermally
calibrated IMU. The IMU chosen is the VN100, produced by Vectornav Co. It
includes a 3-axis accelerometer, 3-axis gyroscope, 3-axis magnetometer,
pressure sensor, temperature sensor, and a 32 bit processor for running an
on-board extended Kalman filter. This IMU has been tested and used in a
series of previous works \citep{7513396,
rabault_sutherland_gundersen_jensen_2017, JGRC:JGRC21649,
marchenko2017field}, which allowed us to both confirm the quality of the data
acquired by the IMU and provide valuable information about wave propagation and
attenuation in landfast ice and some grease ice layers. Using the on-board
Kalman filter together with low-pass filtering of the signal, wave motion
can be accurately measured \citep{doi:10.1175/JTECH-D-16-0219.1}.

The newer version of the instrument, which was tested on landfast ice in
Tempelfjorden, Svalbard in March 2018 and deployed during the Physical
Processes cruise of the Nansen legacy research project in September 2018
\citep{NansenLegacy}, has extended capabilities compared with the simple
logger presented in \cite{rabault_sutherland_gundersen_jensen_2017}. Namely,
it integrates a Raspberry Pi microcomputer which can process the data
in situ and generate compressed spectra from the data recorded, together
with an Iridium modem which enables satellite communications. Moreover, a
low-power unit is added to allow for efficient energy use which, together
with the addition of a solar panel, allows for long term operation. In the
following, the older instruments without in situ processing and satellite
communications will be referred to as ``waves in ice loggers'', as they
basically perform regular 
logging and timestamping of the signal from
the VN100, while the new instruments will be referred to as ``waves in ice
instruments''.

\begin{figure}[h]
  \begin{center}
    \includegraphics[width=.45\textwidth]{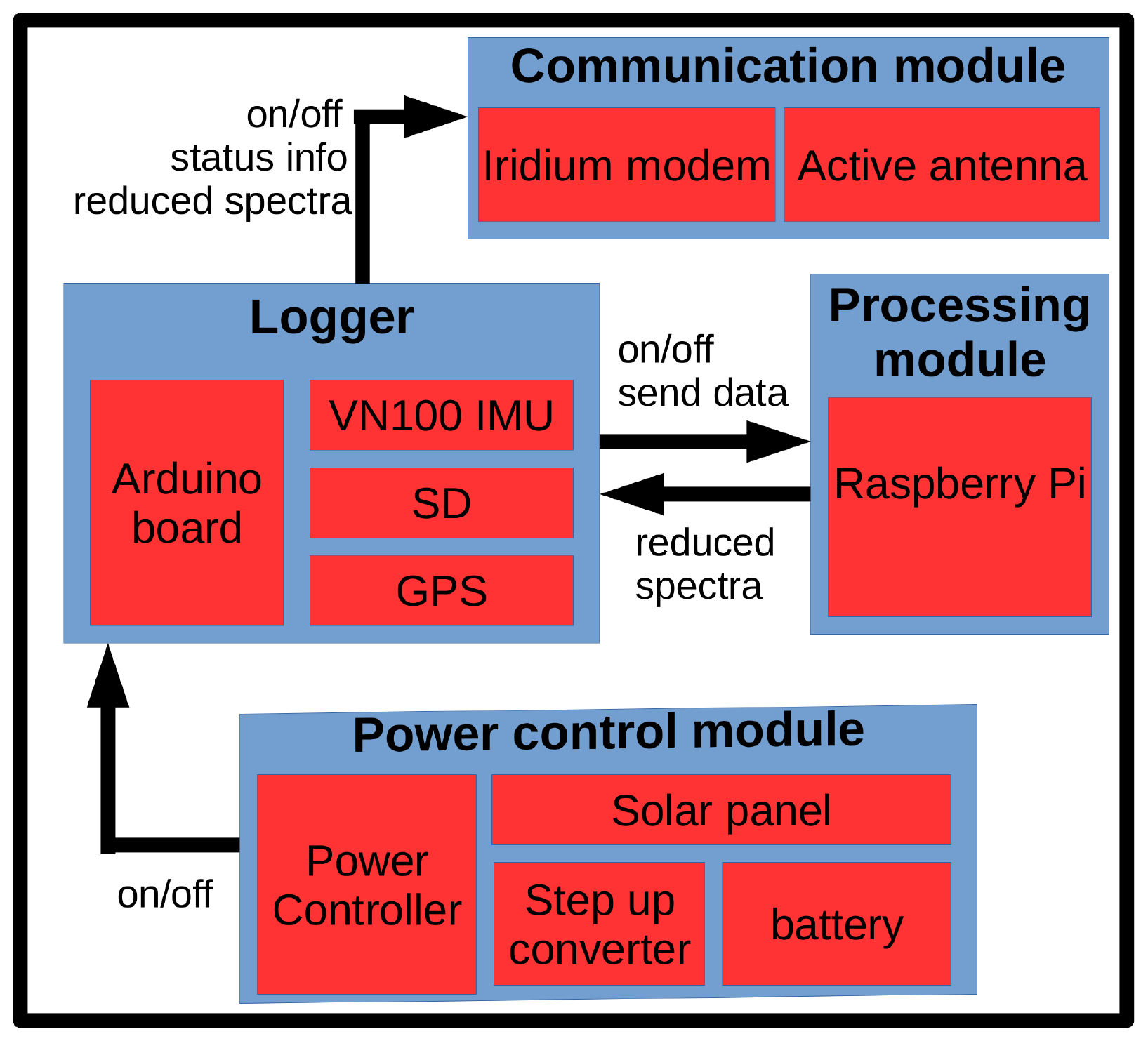}
    \caption{\label{fig_modules} The general architecture of the instrument. Using a modular design, where different components are added on a master Printed Circuit Board (PCB) on top of an Arduino board allows a high level of flexibility for future use.}
  \end{center}
\end{figure}

More specifically, the waves in ice instrument is composed of 4 main modules,
as indicated in Fig. \ref{fig_modules}:

\begin{itemize}
  \item A power control module - composed of a low power microcontroller,
  a LiFePO4 battery cell, a solar panel, and a step-up converter - takes
  care of power management. LiFePO4 battery technology is chosen owing to the
  robustness of the cells, and their ability to withstand low temperatures. The
  step-up converter generates the 5V supply needed by the electronics,
  from the voltage of a single battery cell. The microcontroller takes care
  of implementing the logic of the power control. Namely, it controls the
  charging of the battery by coupling the solar panel, and when the
  logger itself should be waked up. This module is optimized for low energy
  use, being the only one that is always powered on.

  \item The logger itself, which records the wave motion, is composed of
  an Arduino board, the VN100 IMU, a GPS, and a SD card for storing the
  data. It is very similar to the waves in ice logger which was presented in
  \cite{rabault_sutherland_gundersen_jensen_2017}. The logger is activated
  by the power controller approximately every 5 hours, and performs measurements
  of waves in ice for 25 minutes. The use of an Arduino board means that
  other sensors could be easily interfaced in both hardware and software, and
  added to the logics of the instrument. The sleeping time, which decides the
  duration between consecutive measurements, can be modified in the software
  and it is discussed further later in the text.

  \item The processing module, composed of a Raspberry Pi microcomputer
  running a stripped-down version of Linux, takes care of analyzing the
  data generated by the logger, and generating the compressed spectra that
  are sent through Iridium. It communicates with the Arduino board for both
  receiving the waves data, and sending the compressed results.

  \item The communication module, which comprises an Iridium modem and the
  active antenna, allows transmission of data through satellite. The modem
  is driven by the Arduino board. The Short Burst Messages (SBM) protocol,
  allowing messages of 340 bytes to be sent by the modem, and 270 bytes to
  be received from the satellite, is used for communications as it is both
  cost effective and sufficient for sending compressed data.
\end{itemize}

\begin{figure}[h]
  \begin{center}
    \includegraphics[width=.45\textwidth]{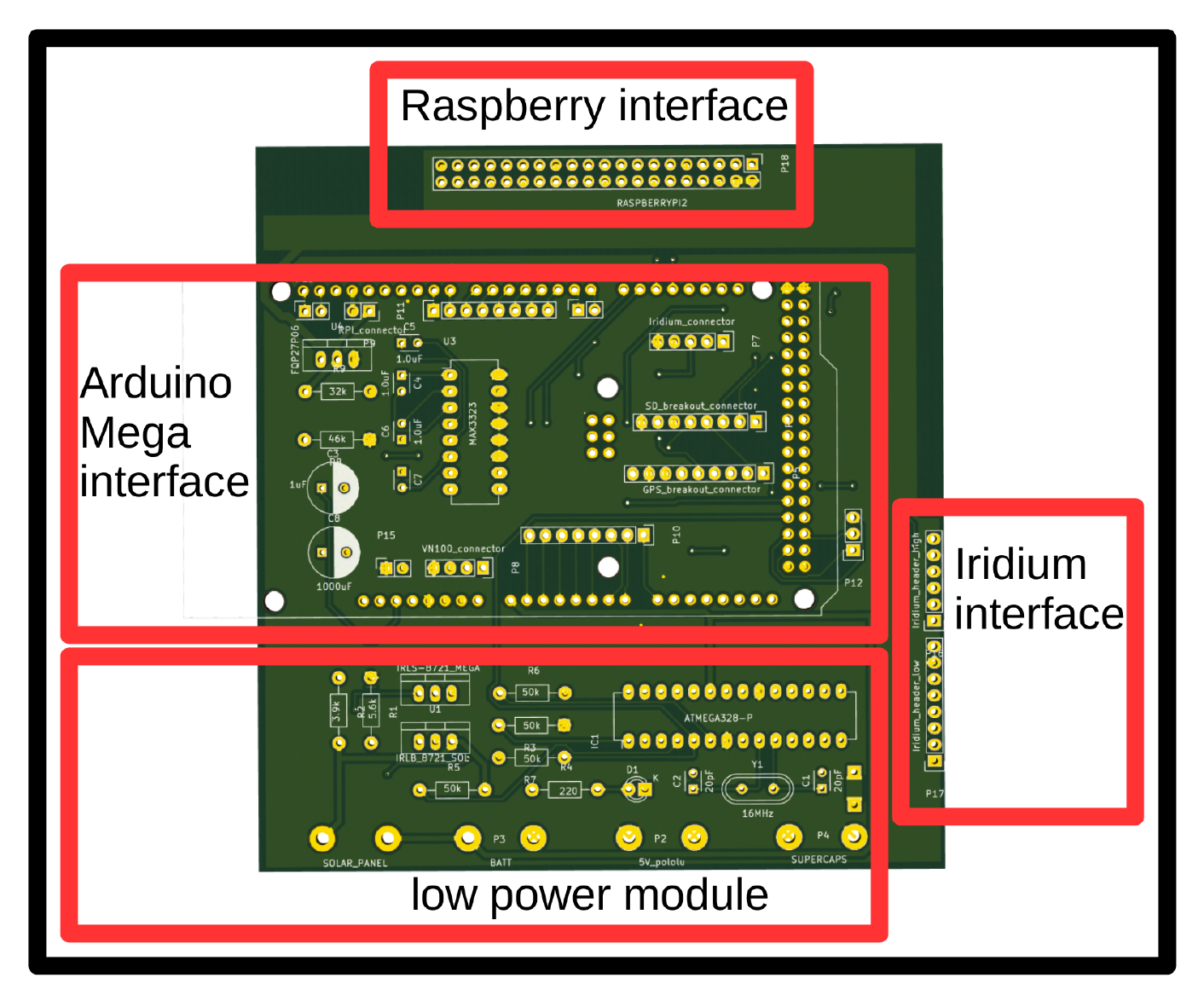}
    \caption{\label{fig_PCB} The general layout of the main PCB on which all components are mounted. The full CAD files are released as open source material, see Appendix A.}
  \end{center}
\end{figure}

All the components are integrated on a central PCB, which connects all the
modules together, see Fig \ref{fig_PCB}. This makes the instrument easy to
produce and assemble. The whole instrument is packaged into a single Pelican
Case with the solar panel mounted on the top, which makes it rugged, compact
and convenient to use in fieldwork (dimensions are $34 \times 30 \times 15$
cm). This packaging of the instrument was chosen as our aim is primarily to
measure waves in ice, and are intended to be deployed
on ice floes rather than floating on water. Of course, the electronics,
sensor and battery could be packaged in another container if the aim would
be to perform measurements while floating in the ocean, in which case the
hydrodynamic response of the instrument would play a role and the hull of the
instrument should be carefully designed. In releasing our design
as open source allows for groups with different needs to easily adapt our
solution to their requirements. In its present form, the complete design
weights about 4.5 kg and all the antennas are mounted inside the case. This
means that the instrument is self-contained, robust and easy to deploy. The
PCB is designed in KiCAD, an open source electronics CAD software and can
be directly produced at a low cost. Therefore, the typical total cost of
the complete instrument is around 2000 US\$, where the IMU itself represents
around 1100 US\$ of the total cost. One instrument can be built in
around 4 to 6 hours of work.

\subsection{Software and in situ data processing}

This section presents the software workflow for workflow the instrument. It should be noted that this workflow can be easily customized in software to change the activity pattern or integrate additional measurements:

\begin{itemize}
  \item At the beginning of each measurement cycle, the low-power unit of the instrument wakes up the logger.
  \item The logger measures waves in ice using the IMU and records GPS information during 25 minutes.
  \item When the measurements are finished, a status message is sent through Iridium. This contains information such as the logger GPS position, the battery level, and some technical information about the logger health.
  \item Following transmission of the status information, the Raspberry Pi (RPi) is triggered to wake up. It receives a copy of the data that has just been recorded, processes the data so as to produce compressed spectra, transmits the data back to the logger, and shuts itself down. The RPi is only awake for a few minutes.
  \item Following shutdown of the RPi, the logger transmits the compressed spectra through Iridium.
  \item Finally, the logger is shut down and only the low-power unit is kept operating. The instrument goes into ultra low power mode for a pre-programmed time interval of around 5 hours, which can be modified in software as previously mentioned.
\end{itemize}

The code for the low power unit and the logger is written in C++, while
the signal processing on the RPi is in Python. The Iridium transmission
cost is kept low as only compressed spectra and some status information are
transmitted. The typical Iridium cost is about 3 \$ a day. While it would be
possible to send more data by transmitting a succession of Short Bust Data
(SBD) messages (according to the data sheet of the Iridium modem, it should be
possible to transmit one SBD packet each 10 seconds), we preferred to avoid
such a solution as this would increase the cost per day (though this is not
the main motivation, as the communication costs remain limited compared with
the price of the instrument), energy consumption (transmitting information by
satellite is quite energy demanding), and most importantly this would require
us to split the data between several SBD packets, which reduces reliability of
the whole system. Indeed, this means that if just one packet as part of a
multi-packet message gets lost, the whole data is potentially corrupted and useless.
Regarding the capabilities of the Iridium modem we use, we also note that
it allows for bi-directional communications with the instrument. This means
that further developments in the software would allow for remotely controlling
the behavior of the instrument, such as the time interval between measurements.
This could allow different groups to adapt our instrument to their needs by
performing minor changes into the software, which is much less time consuming
than developing a new design from scratch. Such updates could also be used
to modify, for example, the frequency at which measurements are performed, so that
even intermittent phenomena could be captured if they are expected to play
an important role \citep{doble2017robust, GRL:GRL52708}.

As a consequence of the limited amount of data that can be transmitted through
Iridium using the SBD packets, the in situ data processing and compression
are important parts of the design. All of the onboard data processing is
performed based on data sampled at 10 Hz. This includes the calculating of
auto- and co-spectra and any integrated spectral parameters we wish to send
by Iridium. In addition, the first five Fourier coefficients, i.e. the 1-D
energy spectrum and the four directional coefficients, are sent via Iridium
at a reduced frequency resolution in order to meet our requirements for
data transmission. Details of the onboard data processing can be found in
the subsequent paragraphs.

The vertical acceleration ($a_z$) and the two orthogonal components of the buoy
slope (i.e. pitch and roll) are recorded internally at 10 Hz on the North,
East and Down reference frame as calculated by the onboard Kalman filter
(running internally at 800Hz together with the raw data acquisition). To
calculate the vertical displacement $\eta$ the vertical acceleration is
integrated twice with respect to time in the frequency domain using the same
method as \cite{ADEVICEKOHOUT}. This is done by calculating the Fourier
transform and using the frequency response weights of $1/\omega^2$ and a
half-cosine taper for the lower frequencies to prevent an abrupt cut-off
\citep{tucker2001waves}, i.e.

\begin{equation}
	\eta(t) = \mathrm{IFFT}\left[H(f)\mathrm{FFT}(a_z)\right],
	\label{eq:elevation}
\end{equation}

\noindent where $\mathrm{FFT}$ and $\mathrm{IFFT}$ denote the Fourier and Inverse Fourier transforms respectively and $H(f)$ is the half-cosine taper function,

\begin{equation}
	H(f) = 
	\begin{cases}
		0, & 0 < f < f_1 \\
		\frac{1}{2}\left[ 1 - \cos{ \left(\pi\frac{f-f_1}{f_2-f_1}\right)\left(\frac{-1}{2\pi f^2}\right) }\right], & f_1 \le f \le f_2 \\
		\frac{-1}{2\pi f^2}, & f_2 < f < f_c, 
	\end{cases}
	\label{eq:response_weight}
\end{equation}
where $f$ is the frequency,  $f_c$ is the Nyquist frequency, and $f_1$
and $f_2$ are the corner frequencies for the half-cosine taper. We use the
same corner frequencies as \cite{ADEVICEKOHOUT} of $f_1 = 0.02 \mathrm{Hz}$
and $f_2 = 0.03 \mathrm{Hz}$, which are suitable for waves of period in the
range of 4 to 20 s. The time series of the vertical displacement is then
the real part of the inverse Fourier transform.

The spectra and co-spectra of these time series can be used to provide the
directional distribution, which is usually written as \citep{kuik1988method}
\begin{equation}
	E(f, \theta) = S(f)D(f, \theta),
	\label{eq:dirspec}
\end{equation}
where $S(f)$ is the 1-D power spectral density (PSD, also referred to as
$PSD_{\eta}$ in the following), calculated from the heave, and $D(f, \theta)$
is the normalized directional distribution which has the property
\begin{equation}
	\int_{-\pi}^{\pi} D(f, \theta) d\theta = 1.
	\label{eq:dnorm}
\end{equation}

While several methods exist to calculate $D(f, \theta)$ \citep{kuik1988method,doi:10.1175/JTECH-D-16-0219.1} they are predominantly based on the following four Fourier coefficients \citep{longuet1961observations}:

\begin{align}
	a_1(f) & = \int_{-\pi}^{\pi} \cos\left(\theta\right) D(f, \theta)d\theta = \frac{Q_{zx}(f)}{k(f)C_{zz}(f)} \label{eq:a1} \\
	b_1(f) & = \int_{-\pi}^{\pi} \sin\left(\theta\right) D(f, \theta)d\theta = \frac{Q_{zy}(f)}{k(f)C_{zz}(f)} \label{eq:b1} \\
	a_2(f) & = \int_{-\pi}^{\pi} \cos\left(2\theta\right) D(f, \theta)d\theta = \frac{C_{xx}(f)-C_{yy}}{k^2(f)C_{zz}(f)} \label{eq:a2} \\
	b_2(f) & = \int_{-\pi}^{\pi} \sin\left(2\theta\right) D(f, \theta)d\theta = \frac{2C_{xy}(f)}{k^2(f)C_{zz}(f)} \label{eq:b2},
\end{align}

\noindent where $C$ represented the auto- and co-spectra and $Q$ the quadspectra, with $x$, $y$ and $z$ denoting the pitch, roll and heave respectively. Here $k(f)$ is the wavenumber, which can be obtained from the known dispersion relation or estimated from the autospectra as:

\begin{equation}
	k(f) = \left(\frac{C_{xx}(f) + C_{yy}(f)}{C_{zz}(f)}\right)^{1/2}.
	\label{eq:k}
\end{equation}

While the software infrastructure necessary for obtaining directional wave information is already
in place, we will not discuss this issue further here. Indeed, more work needs to be done on the current
instrument to make sure that the compass of the IMU is isolated from magnetic disturbances induced by the
battery and electronics.

Equation (\ref{eq:k}) is sometimes used as a quality control flag
\citep{tucker2001waves} when the dispersion relation is known, e.g. with the open water dispersion relation
$k_0 = \omega^2/g$ where $\omega$ is the angular frequency ($2\pi f$) and $g$
is the acceleration due to gravity. For waves in ice, there are several motivations here for
using the open water dispersion relation, rather than a dispersion relation
taking into account the effect of the ice. First, it would be challenging
to perform a direct estimation of the ice thickness from the data recorded,
and such estimation would probably be noisy and brittle, and therefore
unsuited for an autonomous instrument operating on its own. Moreover, it
seems relatively well established that the dispersion relation in thin broken
ice representative of the marginal ice zone (MIZ) is, for practical matters, well described
by the open water dispersion relation as was reported in several studies
\citep{JGRC:JGRC21649, marchenko2017field, doi:10.1175/JPO-D-17-0167.1}. This
is due to the existence of many cracks and regions of open water between the
floes that prevent the transmission of flexural stresses, while at the same
time the added mass effect of the ice is negligible for the typical range of
frequencies encountered. When sending data via Iridium we will use this ratio,

\begin{equation}
	R(f) = \left(\frac{C_{xx}(f) + C_{yy}(f)}{C_{zz}(f)}\right)^{1/2} \frac{g}{\omega^2}.
	\label{eq:G}
\end{equation}

The spectra are calculated using the Welch method~\citep{WelchMethod} and
12000 samples (20 minutes at 10 Hz sampling), using a hanning window on
1024 sample segments and a 50 \% overlap. The Fourier coefficients are then
downsampled into 25 logarithmically equally spaced bins between 0.05 Hz and
0.25 Hz. This type of sampling gives greater resolution at low frequencies (a
minimum resolution of 0.0035 Hz near 0.05 Hz) and less at higher frequencies
(a maximum resolution of 0.0162 Hz near 0.25 Hz). Therefore this downsampling
strategy allows for greater resolution at low frequencies where most of the wave energy
in ice is expected to be prevalent.

All the parameters of this processing algorithm are chosen following
previous work on in situ measurements of waves in ice \citep{kohout2014storm,
ADEVICEKOHOUT}, and can be modified in the software by the user if one needs to
adapt to specific conditions. Of course, modifying some of those parameters,
such as the frequency cutoffs or the frequency range exported, may change the
amount of data generated and may require some additional adaptations of the
compression method. In addition, introducing variability in the configuration
of the instruments may be a source of possible mistakes, and it may be worth
in this case to consider adding some basic information about the frequency
vector of the data transmitted in the SBD messages, to allow cross-checking.

Each Iridium message containing the spectral parameters consists of 340 bytes
(as previously mentioned, this is imposed because of how the Iridium SBD
protocol works). This includes estimates of the significant wave height and
zero-upcrossing, calculated both from the time series as well as from the
spectral moments, as well as the reduced  six spectra: $S(f)$, $a_1(f)$,
$b_1(f)$, $a_2(f)$, $b_2(f)$ and $R(f)$. In order to reduce the number of
bytes sent via Iridium the maximum absolute value ($\mathrm{max}_i$) for
each array is sent as a 32-bit float, and the array is sent as a signed
16-bit integer between $-\mathrm{max}_i$ and $\mathrm{max}_i$.

The significant wave height and zero-upcrossing periods are calculated from
both the full time series and the full spectrum. Both temporal and spectral
methods are used for redundancy checks to compare with the reduced spectrum
also sent via Iridium. The frequency integration bounds used cover the same frequency
range as transmitted by Iridium. The significant wave height can be calculated from
the time series as $H_{St} = 4 \mathrm{std}(\eta)$ where $\eta$ is the
elevation time series. The significant wave height (SWH) can also be calculated
from the spectral moment as $H_{S0} = 4 \sqrt{m_0}$ where the $n_{th}$
spectral moment is defined as
\begin{equation}
	m_n = \int_{0.05}^{0.25} f^n S(f) df,
	\label{eq:moment}
\end{equation}

\noindent where we use the same frequency limits as the ones of the transmitted
spectrum.

The use of integration limits that are more restrained than the complete extent of
the spectra means that $H_{S0}$ is a filtered version of $H_{St}$ (if the whole
spectra were used, then $H_{St}$ and $H_{S0}$ would be equal). Therefore,
comparing those two values can be seen as a simple self-consistency check of the
methodology used.

In the following, we will use the abbreviation ``SWH'' to signify significant
wave height in a general way for example in axis labels, while more detailed
captions will refer to the methodology used by writing either $H_{S0}$
or $H_{St}$.

The typical wave period can also be evaluated in several ways. The
zero-upcrossing period can be calculated from the time series by calculating
the mean time between successive times where $\eta$ goes from positive to
negative, in which case it is referred to as $T_{Z0}$. The zero-upcrossing
period can also be estimated using spectral moments, i.e. $T_{z} =
\sqrt{m_2/m_0}$. In addition, we will refer in the following to the average
crest period $T_c =\sqrt{m_2 / m_4}$, and the spectral peak period $T_p$,
which is defined as the frequency for which the wave spectrum reaches its
maximum. Here also, using different methods to compute the same underlying
physical quantity can be viewed as a simple cross-checking of the data. In
the following, we refer to those quantities in general as the wave period
('WP' in abbreviation), while specific curves will be labeled by the exact
methodology used. A summary of all the variables sent via Iridium, and the
precision of each, can be found in Table~\ref{tab:iridium}.

\begin{table}[h]
\begin{center}
	\begin{tabular}{c|c}
	Variable & Machine precision \\ \hline
        Significant wave height $H_{St}$ (from time series) & 32-bit floating point \\
        Zero-upcrossing period $T_{Z0}$ (from time series) & 32-bit floating point \\
        Significant wave height $H_{S0}$ (from spectral moment) & 32-bit floating point \\
		Zero-upcrossing period $T_z$ (from spectral moments) & 32-bit floating point \\
		magnitude of $S(f)$ & 32-bit floating point \\
		magnitude of $a_1(f)$ & 32-bit floating point \\
		magnitude of $b_1(f)$ & 32-bit floating point \\
		magnitude of $a_2(f)$ & 32-bit floating point \\
		magnitude of $b_2(f)$ & 32-bit floating point \\
		$S(f)$ at 25 frequencies & 25 $\times$ 16-bit signed integer \\
		$a_1(f)$ at 25 frequencies & 25 $\times$ 16-bit signed integer \\
		$b_1(f)$ at 25 frequencies & 25 $\times$ 16-bit signed integer \\
		$a_2(f)$ at 25 frequencies & 25 $\times$ 16-bit signed integer \\
		$b_2(f)$ at 25 frequencies & 25 $\times$ 16-bit signed integer \\
		$R(f)$ at 25 frequencies & 25 $\times$ 16-bit signed integer 
	\end{tabular}
	\end{center}
	\caption{Transmitted data via Iridium. A total of 340 bytes is sent per Iridium message.}
	\label{tab:iridium}
\end{table}

\begin{figure}[h]
  \begin{center}
    \includegraphics[width=.45\textwidth]{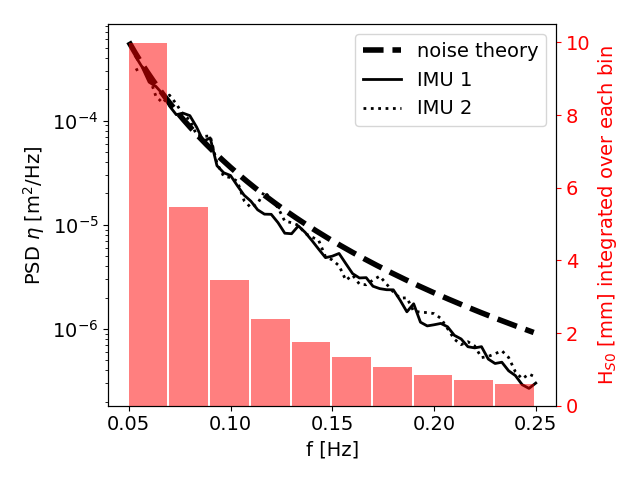}
      \caption{\label{fig:denoising}Illustration of the stable noise level of the IMU after processing with the Welch algorithm. The IMUs 1 and 2 are two different VN100 IMUs that were used for tests on still ground. The noise level is consistent across both IMUs, which can be used to denoise afterhand the transmitted spectra. The noise theory is obtained by fitting a theoretical noise shape to the data on the high frequency part of the spectra, where the effect of noise is most important due to the double integration performed to compute the wave elevation. Finally, the histogram presents $H_{S0}$ obtained by integrating over the width of each of the bins, and provides an estimate of the effective noise level of the instruments at each frequency.}
  \end{center}
\end{figure}

After receiving the compressed data via Iridium, similar data decompression is implemented and
applied to decrypt the messages
transmitted. Moreover, a step of de-noising is applied to the transmitted
spectra on the receiver side. Applying denoising on the receiver side allows
to switch it on and off, to perform quality checks if necessary. Namely, the
IMU has a stable noise characteristic which translates into a reproducible
background noise on the spectra as illustrated in Fig. \ref{fig:denoising}. In
all the following, denoising will only be discussed and applied to the 1-D PSD
(i.e., $S(f)$), but this approach could be generalized to the other Fourier
coefficients that are being transmitted.

While some information about spectral noise levels are available in the
datasheet of the IMU, this is often difficult to translate into a real-world
estimate of the noise background due to the internal Kalman filtering, and
processing applied on the signal. Therefore, the noise level as a function
of frequency $n(f)$ is obtained by fitting a theoretical shape of the form
$n(f) = (9.81 \times 10^{-3} C)^2 (2 \pi f)^{-4}$ to a record obtained on
still ground, where $C$ has unit mg$/\sqrt{\mathrm{Hz}}$ (where we mean ``milli g'' by mg, and the renormalization numerical constant
of value $9.81 \times 10^{-3}$ has the unit of ms$^{-2}$/mg to enforce dimensionality) and characterizes
the noise level. This specific noise shape is chosen as it describes the
effect of a uniform spectral noise density on the acceleration measurements
when converted into the elevation wave spectra, taking into account the double integration
that takes place during the processing. As visible in Fig. \ref{fig:denoising},
this theoretical noise shape is well verified. We find experimentally that
$C=0.24$mg$/\sqrt{\mathrm{Hz}}$ describes the obtained noised satisfactorily,
which compares reasonably well with the product specification of the
VN100 (which states that the spectral noise density should be $0.14$
mg/${\sqrt{\mathrm{Hz}}}$, \cite{VN100specs_2}), taking into account that
the datasheet value may be an optimistic estimate obtained in the laboratory.

In practice, the effect of noise can be neglected for waves larger than
around 10 mm at a frequency of 0.15Hz. However, in the case when small waves
(i.e. typically under 5 mm at 0.15Hz) are observed, correcting for this
background noise improves the quality of the measurements especially for the
lowest frequency where the noise introduced by the double integration of the
vertical acceleration is largest. More specifically, we consider that the
wave signal $s(t)$ and the IMU noise $n(t)$ are uncorrelated, and we want
to compute the power spectral density of the combined output $w(t) = s(t) +
n(t)$. Remembering that the power spectral density is the Fourier transform
of the auto-correlation function $\Gamma$, we mean to calculate:

\begin{equation}
    \Gamma_{w, w}(\tau) = \mathbb{E}\left[ (s(t) + n(t))(s(t + \tau) + n(t + \tau)) \right].
\end{equation}

Applying the distributivity of the expected value operator and using the
condition that the signals $s$ and $n$ are uncorrelated, the cross-terms
disappear leading to $\Gamma_{w, w}(\tau) = \Gamma_{s, s}(\tau) + \Gamma_{n,
n}(\tau)$, therefore $PSD(w) = PSD(s) + PSD(n)$. This means that the stable
sensor noise level can be subtracted from the power spectral density of the
transmitted spectra to reduce noise. In the following, this processing will
be applied to the wave elevation spectra.

\section{Deployment on the ice}

\subsection{Deployment on landfast ice in Tempelfjorden, Svalbard and
validation of the on-board algorithms}

In this subsection, we present a cross-validation of the algorithms
and methodology used for the on-board processing and Iridium data compression
by comparing results obtained from the raw data recorded by a waves in
ice logger and processed using the methodology previously presented in
\cite{rabault_sutherland_gundersen_jensen_2017}, against the results obtained
from a waves in ice instrument, which were transmitted over Iridium.

\begin{figure}[h]
  \begin{center}
    \includegraphics[width=.23\textwidth]{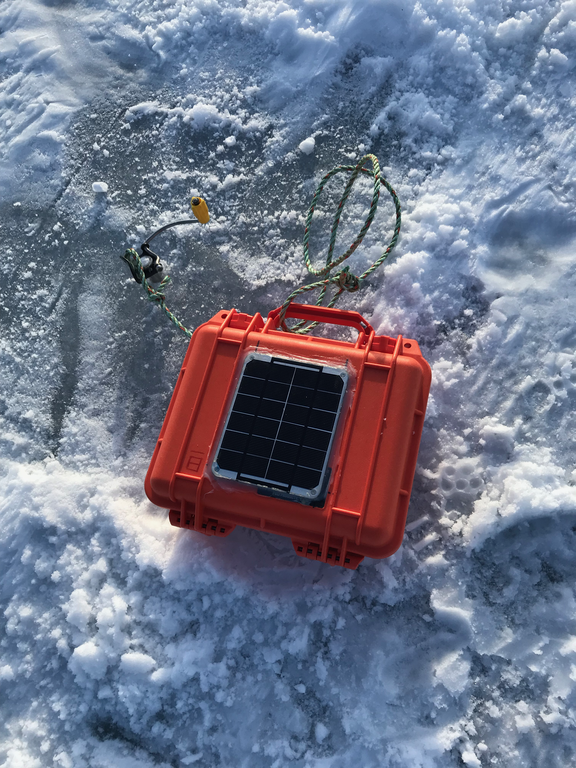} 
    \includegraphics[width=.23\textwidth]{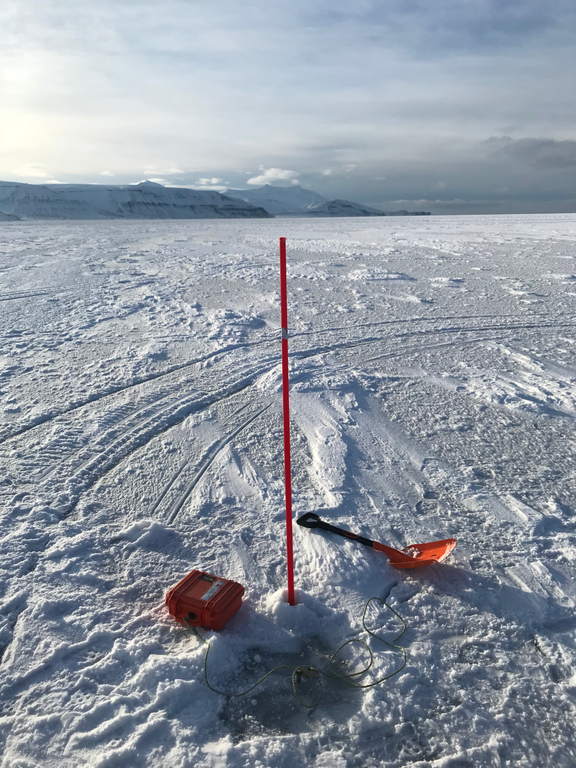}
      \caption{\label{pct_deployment_Svlbd} Illustration of the deployment of a waves in ice instrument on landfast ice in Tempelfjorden, Svalbard (left), and waves in ice logger deployed on its side (right). The shovel gives an idea of the size of the instrument boxes.}
  \end{center}
\end{figure}

The data were obtained during a deployment in landfast ice in Tempelfjord,
Svalbard performed in March 2018. The ice conditions were similar to what were
encountered in an earlier deployment in 2015 \citep{JGRC:JGRC21649}, see Fig
\ref{pct_deployment_Svlbd}. A waves in ice logger and a waves in ice instrument
were deployed side-by-side on the frozen fjord, around 500 meters from the
ice edge. Both performed measurements from March 21st to March 27th. Surface waves were observed 
between the evening on March 22nd and early afternoon on
March 23rd. As the deployment took place on landfast ice in the inner part
of a fjord, the waves were small (the peak significant wave height was about
2.5 cm), but nonetheless could be reliably measured by our instruments.

\begin{figure*}[h]
  \begin{center}
    \includegraphics[width=.69\textwidth]{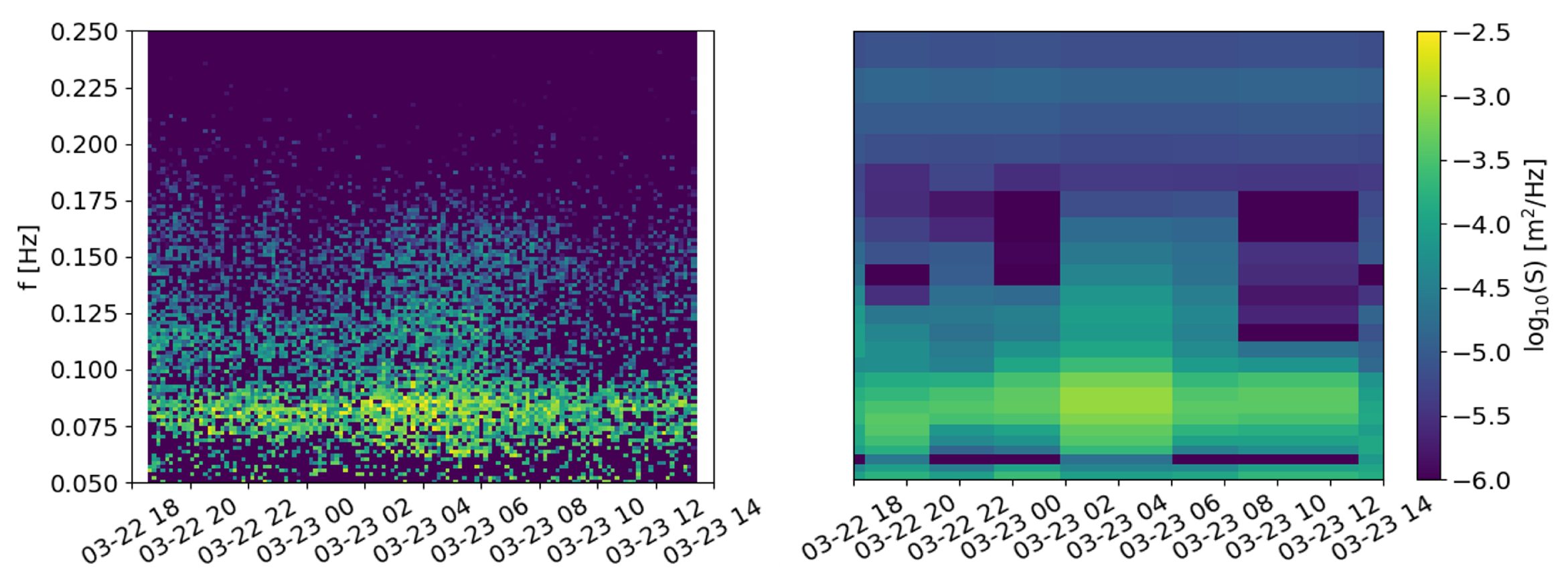}
    \caption{\label{fig_comparison_onboard_iridium} Comparison between the
    spectrogram produced from analyzing the data logged continuously by a
    waves in ice logger (left), and the spectra produced in situ and transmitted using Iridium by a waves in ice instrument (right). The data were generated during a deployment on landfast ice in Tempelfjorden, Svalbard in March 2018. While the frequency and time resolutions are, obviously, lesser in the second case, both the energy level and frequency distribution of the wave signal are very similar in both cases.}
  \end{center}
\end{figure*}

The spectrograms obtained by both instruments are presented in
Fig. \ref{fig_comparison_onboard_iridium}. Obviously, the resolution in time
is much higher in the case of the logger (which records data continuously)
than with the iridium-enabled instrument which transmits data only around
each 3 hours (as this was a test run over a shorter amount of time,
the wakeup frequency was increased compared to the standard value of 5
hours). The resolution in frequency is also reduced, due to the integration
and down-sampling of the spectra, which has the effect of reducing the resolution of
the individual spectra. However, both the distribution and typical value of
the power spectral density are satisfactorily reproduced by the under-sampled
spectrogram (this will also be shown in more details in the next paragraph). In
addition, we also want to stress that, as was mentioned previously,
the wakeup and transmission frequency could easily be increased if deemed
necessary for a different task, and it would be even possible to control this
parameter through the 2-ways iridium link by developing additional software.

\begin{figure*}[h]
  \begin{center}
    \subfloat[2018-03-22 21:00]{\includegraphics[width=0.32\textwidth]{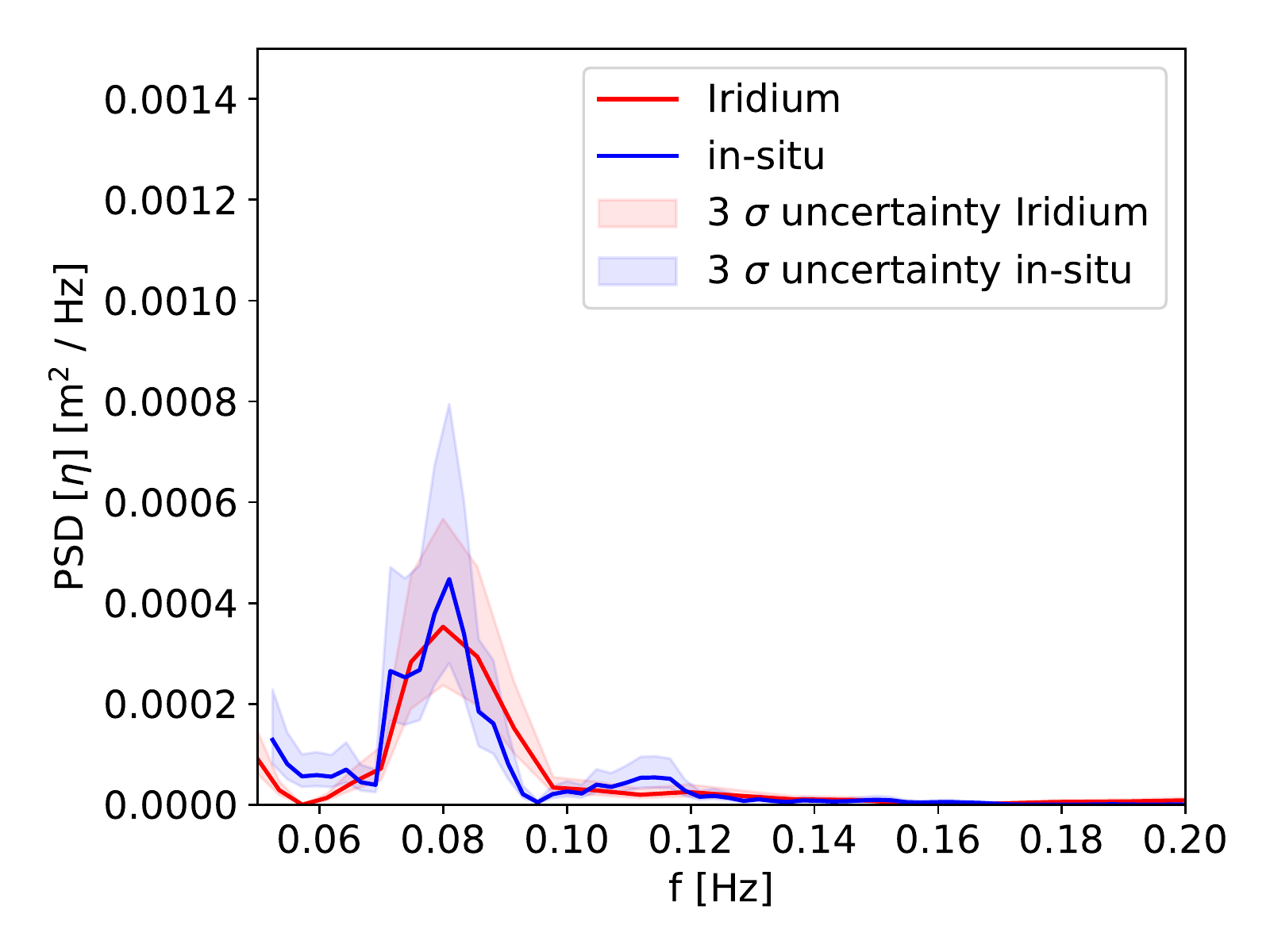}}
    \subfloat[2018-03-23 02:00]{\includegraphics[width=0.32\textwidth]{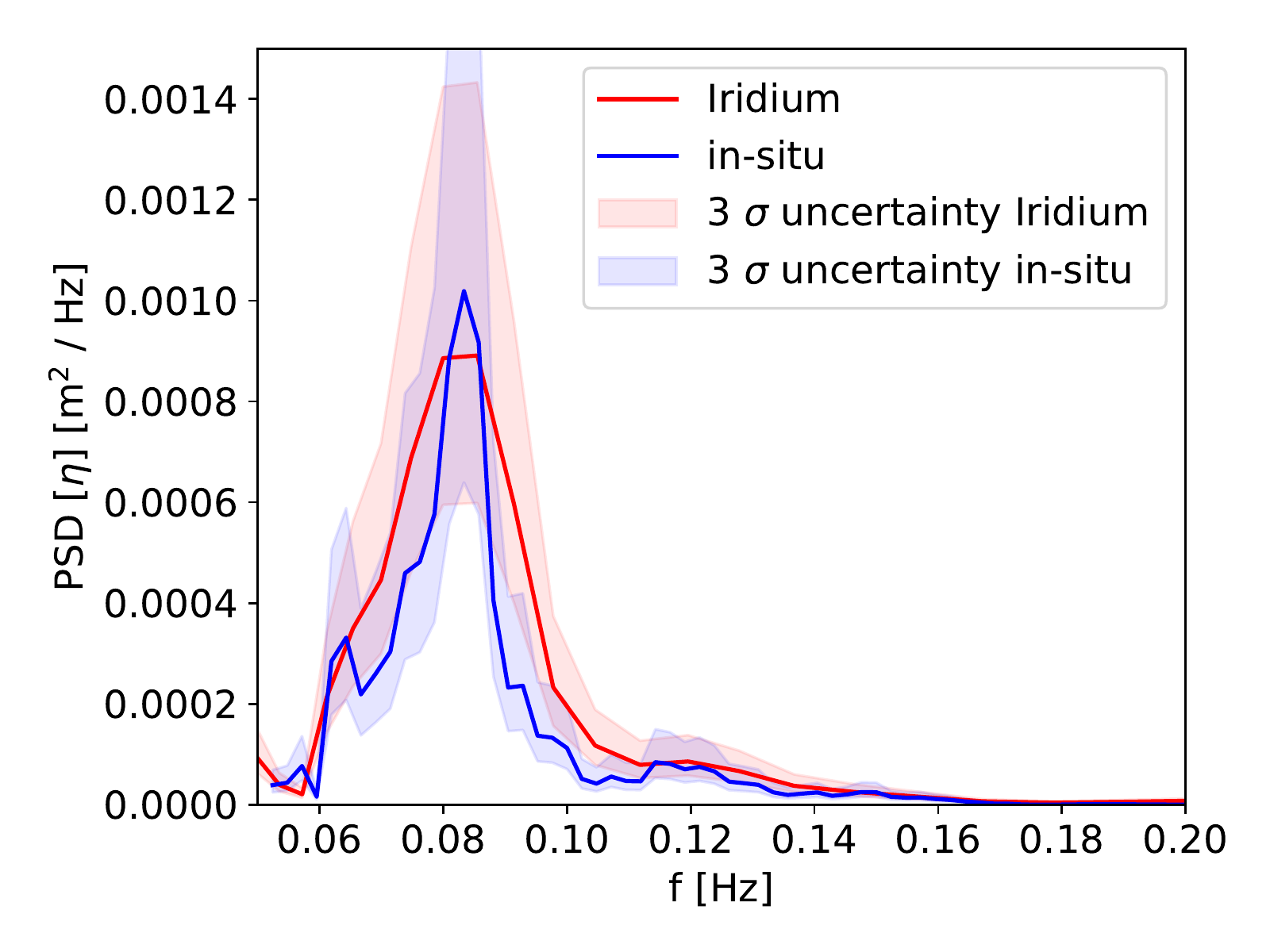}}
    \subfloat[2018-03-23 18:00]{\includegraphics[width=0.32\textwidth]{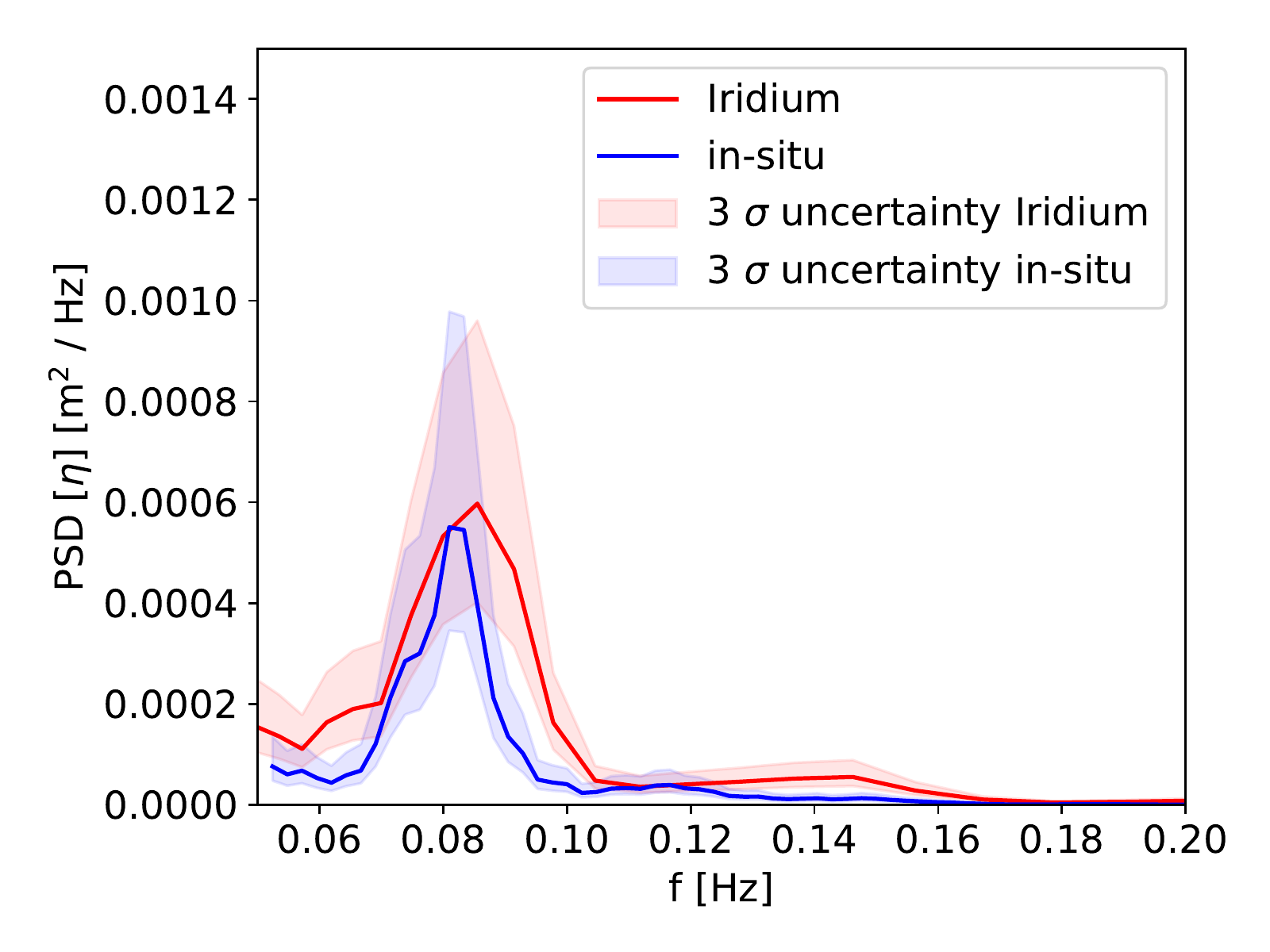}}
      \caption{\label{fig_at_peak_SWH} Comparison between the wave spectra obtained from the waves in ice instrument and the waves in ice logger before, during, and after the peak wave activity recorded. This provides a cross-validation of the processing algorithm and of the data compression methodology used in the waves in ice instrument.}
  \end{center}
\end{figure*}

A more detailed comparison between the results obtained by the
waves in ice instrument and the waves in ice logger is presented in
Fig. \ref{fig_at_peak_SWH}. As visible in Fig. \ref{fig_at_peak_SWH},
the spectra agree well, which provides a more detailed illustration
of the results presented in Fig. \ref{fig_comparison_onboard_iridium}. In
this figure, and in all the following when error bars will be presented
either for wave spectra or for reports of significant wave height, a
3-$\sigma$ confidence interval is used. Confidence intervals are calculated
from the Chi-squared distribution, following the methodology presented
in \citet{doi:10.1061/(ASCE)0733-950X(1986)112:2(338), YOUNG1995669,
TechReportChiSpectra}, where the number of degrees of freedom is calculated
from the number of overlapping segments used in the Welch method \citep{ADEVICEKOHOUT}.

Therefore, the agreement observed in this subsection cross-validates the in situ, on board
data processing against the data processing our group has been using in the past, as well as the data compression and decompression algorithms
used to transmit the reduced spectra over Iridium. This is especially
important as there are many fully autonomous steps involved in the on-board
data processing, data compression and transmission, and therefore the results of this
section offer a cross-validation suggesting that the algorithms and implementations
used are indeed functioning correctly in the autonomous instrument when no human
expertise or inspection is available.

\subsection{Deployment on a large ice floe in the Marginal Ice Zone in the Barents sea, and comparison with the data obtained from nearby pressure sensors}

In this subsection, we compare the results obtained using a waves in ice
logger with those of a nearby pressure sensor mounted under the ice during a
field campaign on a large ice floe in the Barents sea during May 2016. Both datasets
have been presented previously in the literature \citep{marchenko2017field},
and here we simply offer a more detailed assessment of the agreement between
the two methodologies as a way to validate our instrument. As the waves
in ice logger has been cross-validated against the waves in ice instrument
in the previous subsection, this constitutes a validation of both of them
against a different measurement technique.

In all of the following, the data from the IMU loggers will be analyzed
following the same methodology as in the previous sections, while the
analysis of the pressure records will follow the methodology presented in
\citep{MarchenkoMSIFS, marchenko2017field}. The pressure measurements were
performed using a Sea Bird SBE39 plus equipped with a laboratory-calibrated
pressure probe, measuring at a sampling rate of 2.0Hz. The relation in the
spectral domain between the pressure and elevation spectra can be written
as \citep{MarchenkoMSIFS, marchenko2017field}:

\begin{equation}
    S_p(z, f) =  S_{\eta}(f) \left[ P(k, z) \rho_{w} g \right] ^2,
\end{equation}
with $P(k, z)$ a transfer function defined as:
\begin{equation}
    P(k, z) = \frac{\cosh(k(z+H))}{\cosh(kH)} - 1,
\end{equation}
where $S_p(z, f)$ is the spectrum of the pressure fluctuations at depth $z$
(in our case, $z=11$m), $\rho_w$ is the density of water, $g$ the acceleration
of gravity, $S_{\eta}(f)$ the wave elevation spectrum, $k$ the wavenumber,
and $H=160$m the water depth. As measured in \cite{marchenko2017field},
the dispersion relation is not affected by the relatively thin ice floe
in the frequency band where wave motion is present, and therefore we use
the dispersion relation for open water of intermediate depth, $\omega^2 =
g k \tanh(k H)$, with $\omega = 2 \pi f$ the angular frequency.

\begin{figure*}[h]
  \begin{center}
    \includegraphics[width=.45\textwidth]{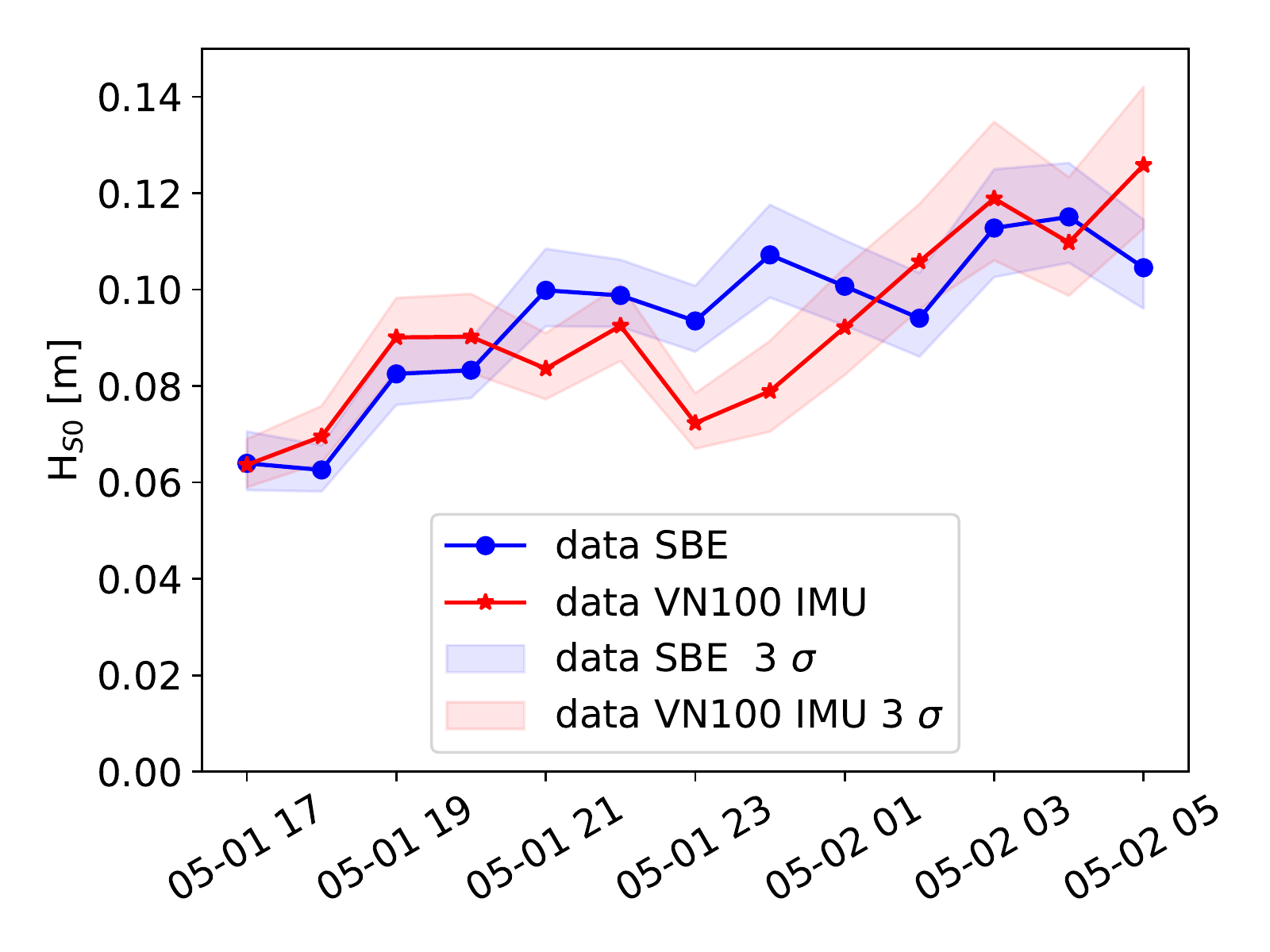}
    \caption{\label{fig_comparison_pressure_imu_SWH} Comparison between the significant wave height computed from the IMU-based and pressure-based instruments, as a function of time. The colored area presents the confidence intervals at 3-$\sigma$. Both the general trend and the typical value of the significant wave height are in good agreement between the two methodologies.}
  \end{center}
\end{figure*}

\begin{figure*}[h]
  \begin{center}
    \includegraphics[width=.45\textwidth]{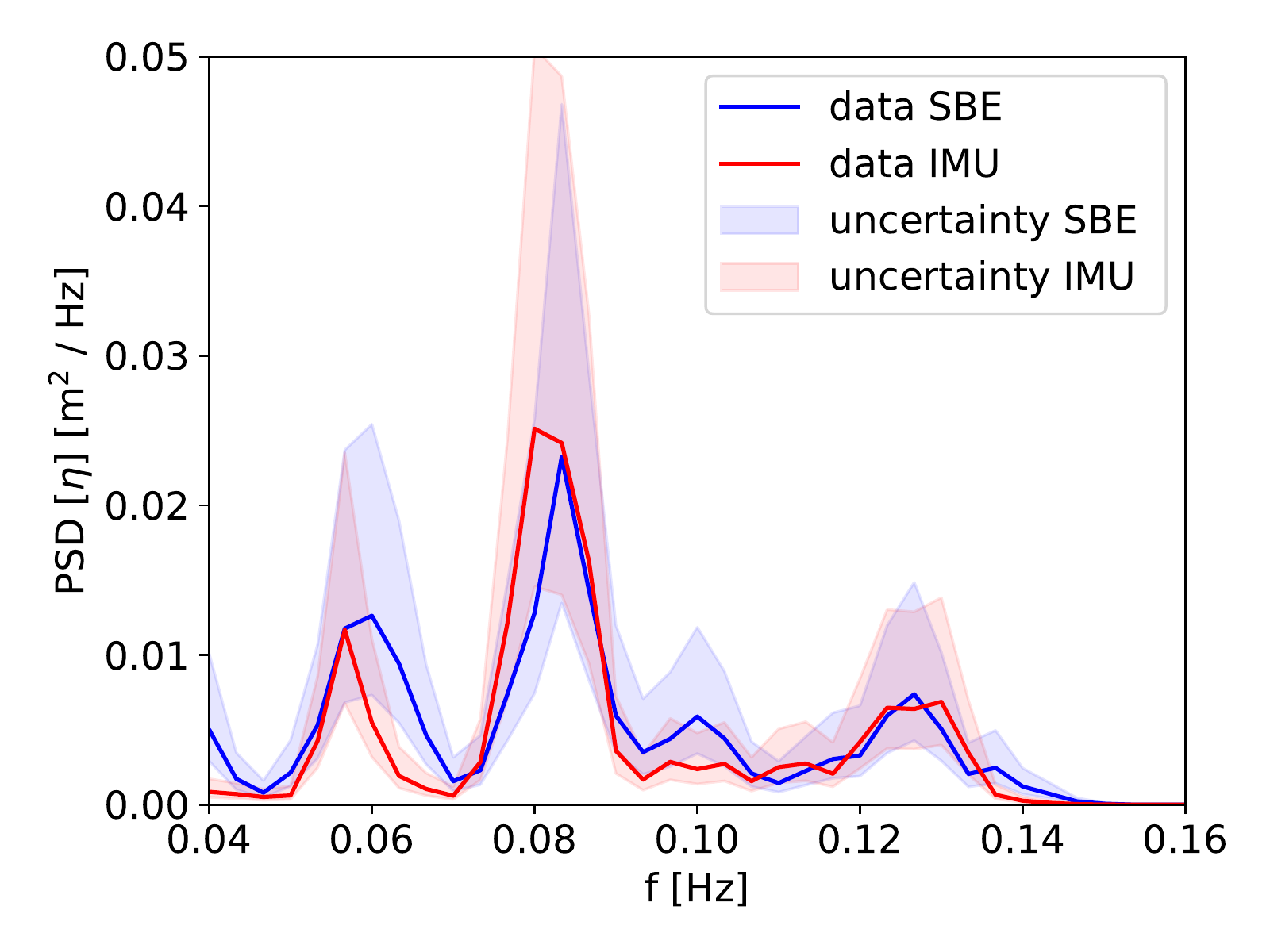}
    \includegraphics[width=.45\textwidth]{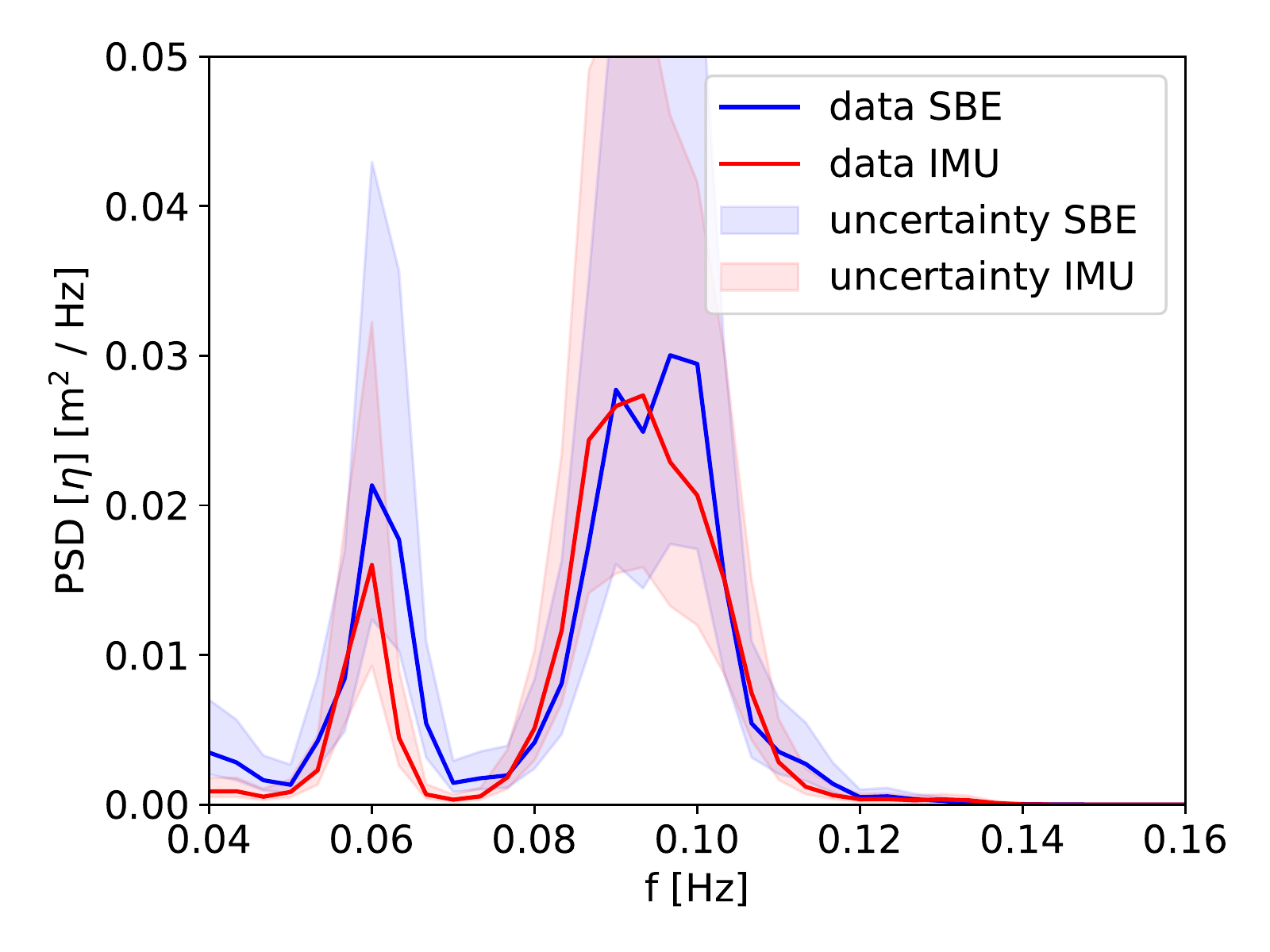}
      \caption{\label{fig_comparison_pressure_IMU_spectra} Comparison between the frequency spectra for the wave elevation computed based on the pressure (SBE) and acceleration (IMU) data. The colored area indicates the 3-$\sigma$ confidence intervals. The spectra on the left are taken on May 1st 2016 at 22.00.00 UTC, while the spectra on the right are taken on May 2nd 2016 at 06.00.00 UTC. Excellent agreement is obtained regarding both the location of the spectral maxima and their amplitudes.}
  \end{center}
\end{figure*}

Following this methodology, a typical peak significant wave height of about
$12$ cm is obtained with both the IMU-based logger and the pressure sensor, as
visible in Fig. \ref{fig_comparison_pressure_imu_SWH} and already reported in
\cite{marchenko2017field}. In addition, both the evolution in time of the
significant wave height and sample spectra taken at specific times (presented
in Fig. \ref{fig_comparison_pressure_IMU_spectra}) demonstrate good agreement
between the two measurement techniques. Although a slight difference (at the
limit of statistical significance) can be observed between the waves in ice
instrument and the pressure sensor for frequencies of around 0.7 Hz, we believe
that this can come from either oscillations of the pressure sensor (the
pressure sensor is attached under the ice on a weighted rope, and therefore
small oscillations of the sensor may be introduced by the waves and ice
motions), and / or from the fact that both instruments were deployed a few
meters away from each other, therefore, the response of the ice floe to the
waves may be slightly different at both locations. In addition, pressure sensors
may also measure pressure fluctuations that are appearing due to local hydrodynamic
processes in the water, for example originating from waves / current interaction
or eddy processes, that may add some energy in specific frequency ranges. Therefore, we consider that
the data provided in this section are in good enough agreement that they successfully validate our results against a
different instrument and measurement technique.

\subsection{Deployment in the Marginal Ice Zone in the Northeast Barents sea, real-world testing of the instruments, and comparison with data from a waves model}

\begin{figure*}
  \begin{center}
    \includegraphics[width=.24\textwidth, height=.11\textheight]{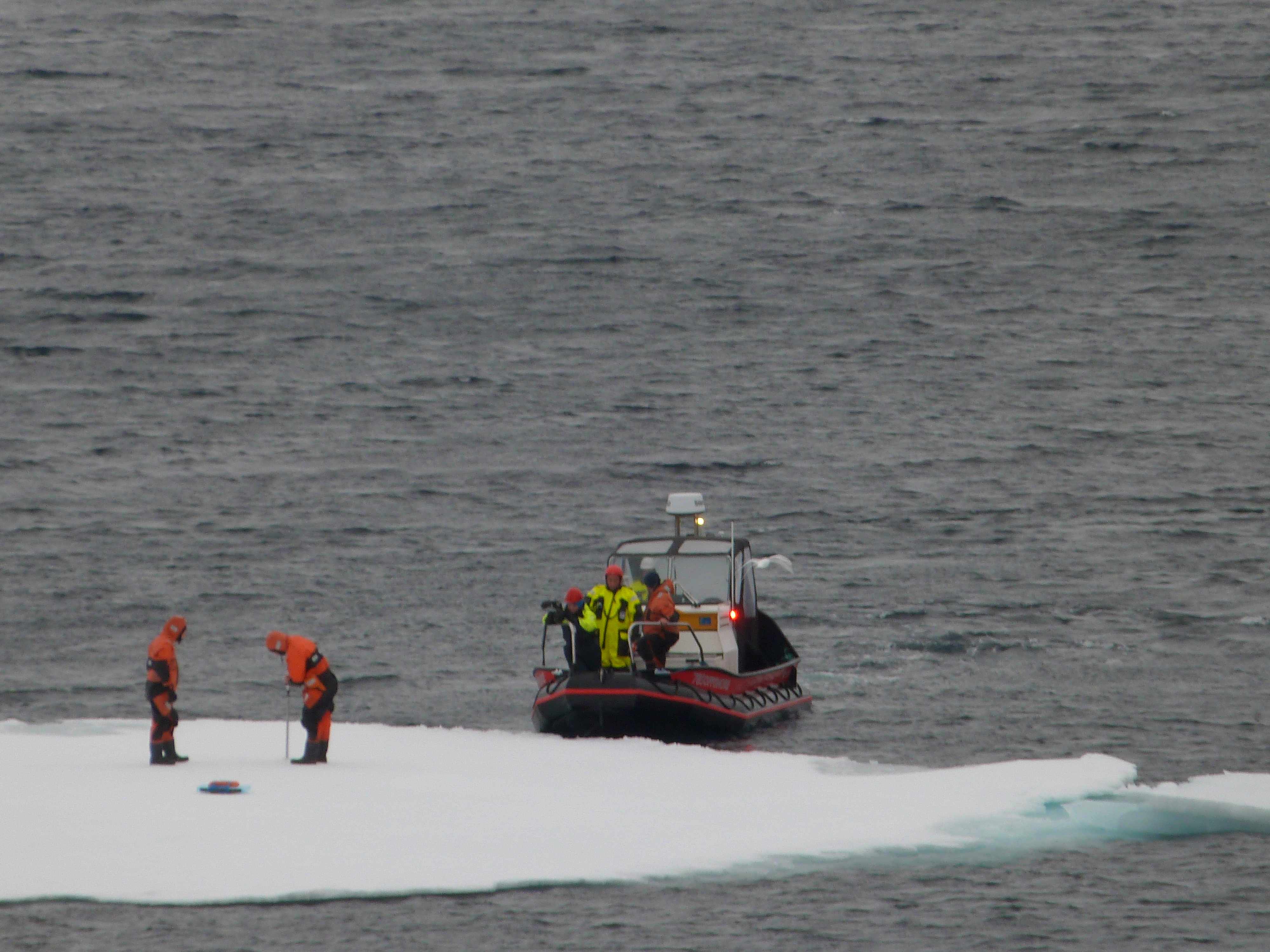}
    \includegraphics[width=.24\textwidth, height=.11\textheight]{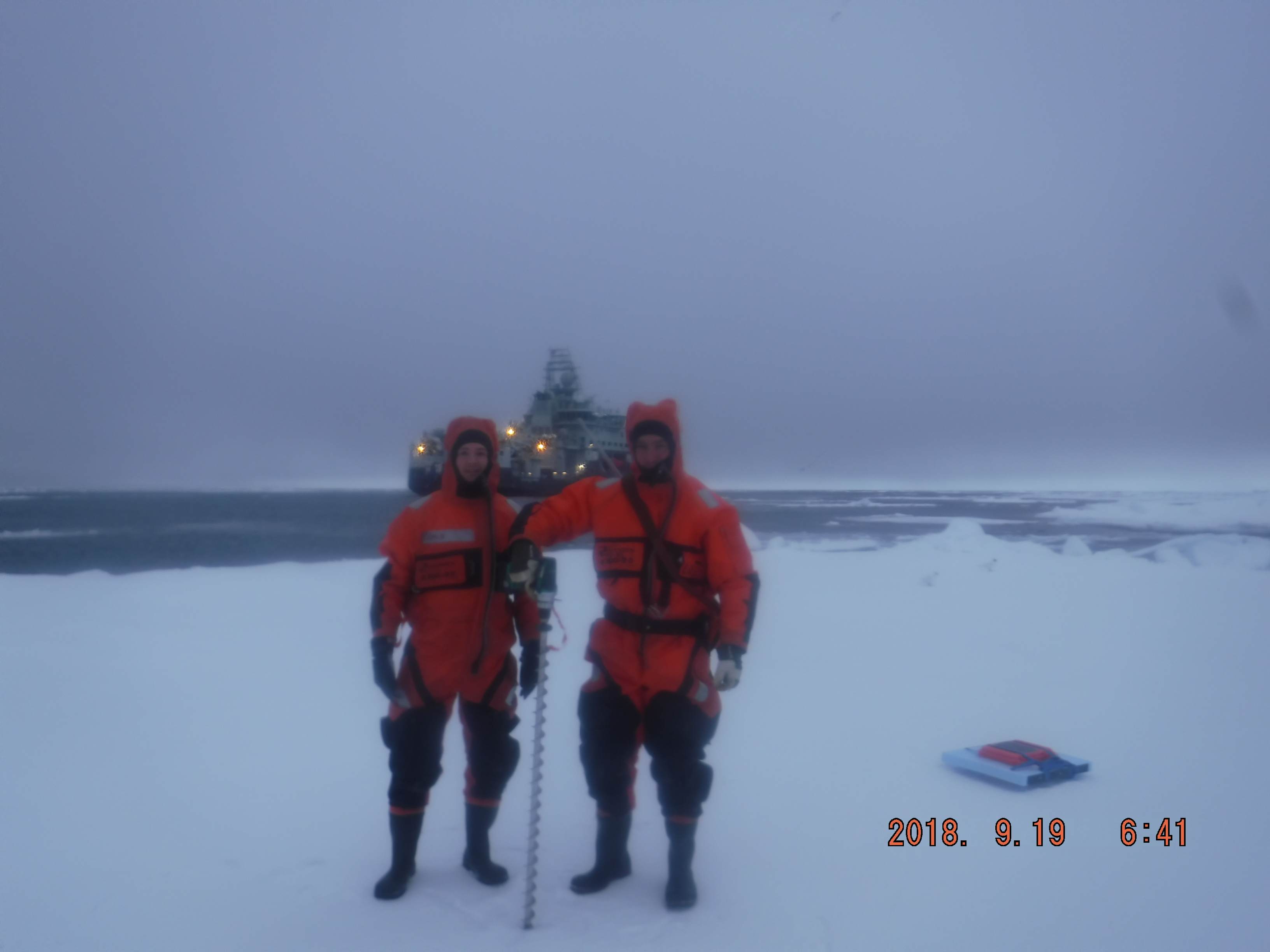}
    \includegraphics[width=.24\textwidth, height=.11\textheight]{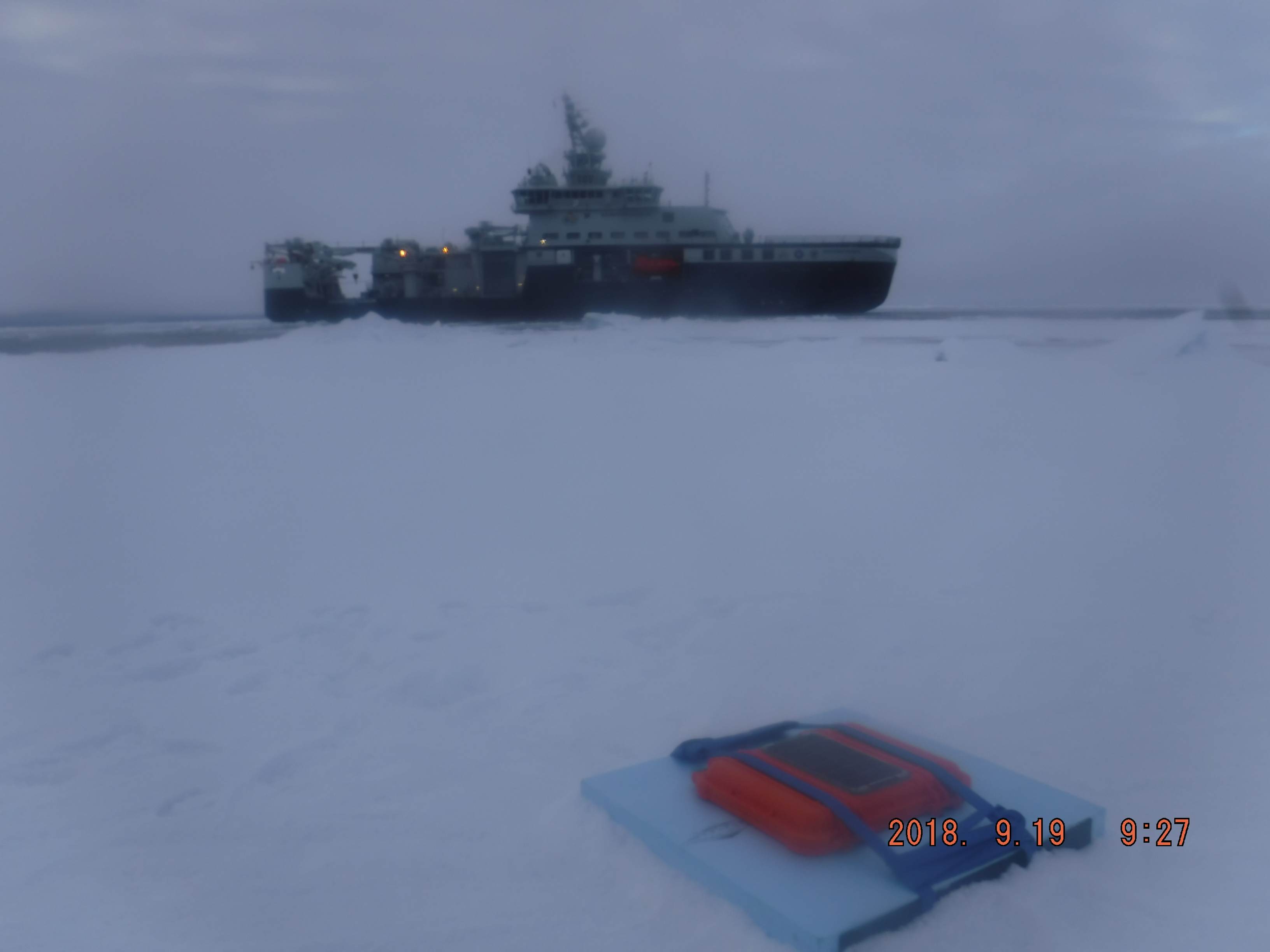}
    \includegraphics[width=.24\textwidth, height=.11\textheight]{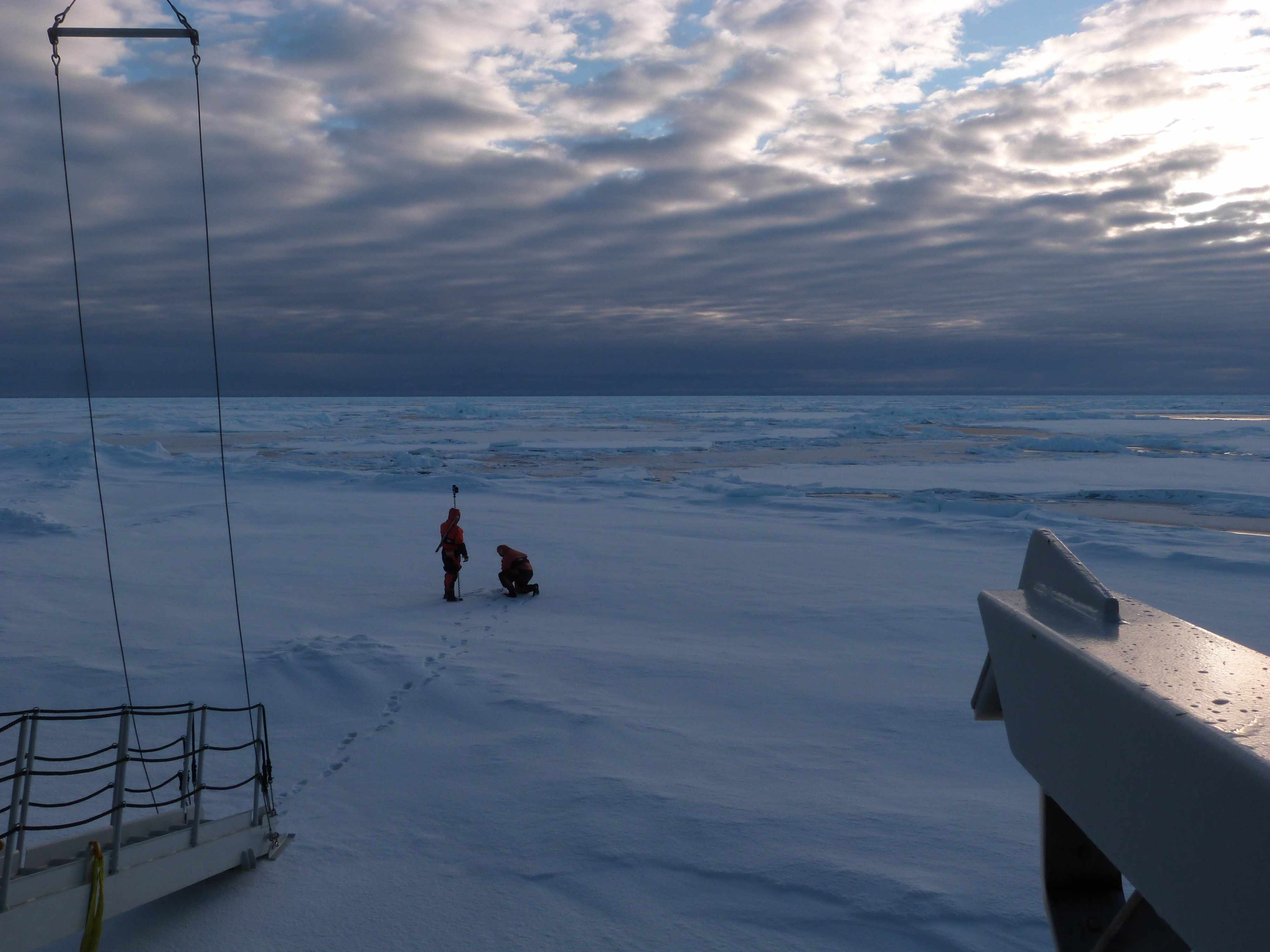}
    \caption{\label{fig_deployment} Deployment of the instruments 1 to 4 (left to right) in the Marginal Ice Zone, Northeast Barents sea. The concentration of ice increases from instrument 1 to 4, as visible in the pictures (1/10th, 3/10th, 9/10th, 10/10th). The instruments, equipped with buoys, are visible on the ice. Photos: credit Malte M\"uller and Lars R. Hole, Norwegian Meteorological Institute.}
  \end{center}
\end{figure*}

In this subsection, we present early results obtained in the course of the
Nansen legacy cruise which took place in September 2018. The aim of this
subsection is not to present a detailed scientific analysis of the results,
which will be performed in future work, but to check and validate the proper
functionality of the waves in ice instruments regarding energy consumption, GPS
tracking function, and 1D wave spectrum estimates. A total of 4 instruments
(named 1 to 4 in the following) were deployed from the research vessel
RV Kronprins Haakon on September 19th. The deployment took place in the
MIZ, Nortwest of Svalbard. During the deployment, the
vessel was steaming inside the MIZ and the pack ice. As a consequence,
the 4 instruments were deployed as an array with the first sensor being
located on an ice floe in the outer MIZ (ice concentration 1/10th), the
second on an ice floe further in the MIZ (ice concentration 3/10th), the
third at the beginning of the closed pack ice (ice concentration 9/10th),
and the fourth inside the closed pack ice (ice concentration 10/10th), as
illustrated in Fig. \ref{fig_deployment}. During deployment, the instruments
were equipped with a floating device and buried half-way inside the snow,
with only the top of the case (hosting the solar panel and the antennas)
directly pointing to the sky.

\begin{figure}
  \begin{center}
    \includegraphics[width=.45\textwidth]{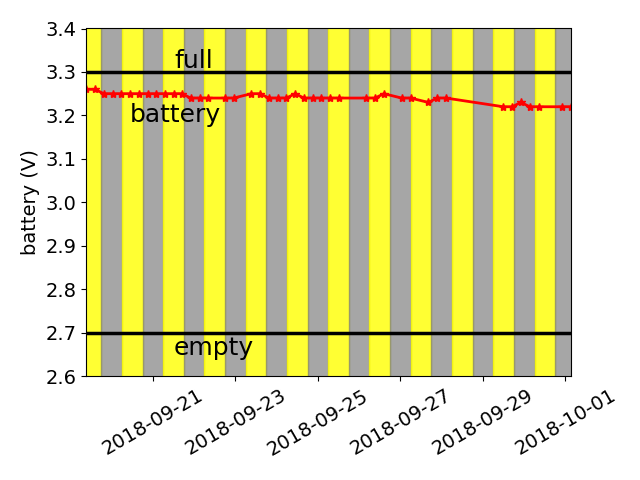}
    \caption{\label{fig_batt} Time evolution of the battery level of instrument 2. The background color indicates time of the day: yellow is for 06:00 to 18:00, gray is for 18:00 to 06:00. The full and empty battery levels correspond to $3.3$V and $2.7$V, respectively. The battery level remains very high thanks to efficient power management and the presence of a solar panel.}
  \end{center}
\end{figure}

The temporal evolution of the battery level of the second instrument, which was the one which transmitted for the longest time, is presented in Fig. \ref{fig_batt}. As previously stated, the battery technology used inside the instruments is LiFePO4, and according to the datasheet the voltage of a fully charged battery is around 3.3V, and a fully discharged battery has a voltage of around 2.7V. As visible in Fig. \ref{fig_batt}, the battery barely depleted over the period of 12 days for which it was deployed. This is due to both the high power efficiency of the instrument, and the presence of the solar panel that provides energy at least for the first days of deployment, until the start of the polar night. It should be noted that temperature can also influence the battery voltage, which explains for apparent increases in battery level during some night periods. It is also apparent from Fig. \ref{fig_batt} that the battery level was not the cause for the end of transmission. This confirms the quality of the power management solution and overall design, and is further discussed later in this section.

\begin{figure}
  \begin{center}
    \includegraphics[width=.35\textwidth]{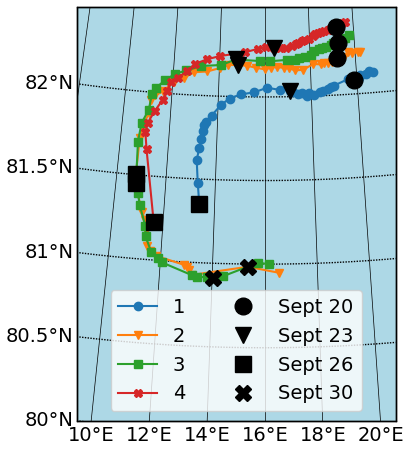}
    \caption{\label{fig_track} The drift of the instruments, as obtained from the GPS strings transmitted through Iridium. The symbols indicate the position of each instrument for the first transmission received on the corresponding day. A general drift pattern is clearly visible.}
  \end{center}
\end{figure}

The GPS information transmitted by all 4 instruments is presented in Fig. \ref{fig_track}. As visible in Fig. \ref{fig_track}, the trajectories of the instruments are well resolved until transmissions are lost. While some dropouts are present, a pattern of drift first to the West, then South and South East is clearly visible. During this period of 12 days, the instrument 2 (which survived the longest) drifted approximately $340$ km. This corresponds to the effect of the transpolar current, together with the forcing created by a storm present in the region September 24th and 25th. The data will be used for validating satellite tracking of ice drift and models, and these results also indicate that the instruments are valuable to use as drifters.

\begin{figure*}
  \begin{center}
    \includegraphics[width=.65\textwidth]{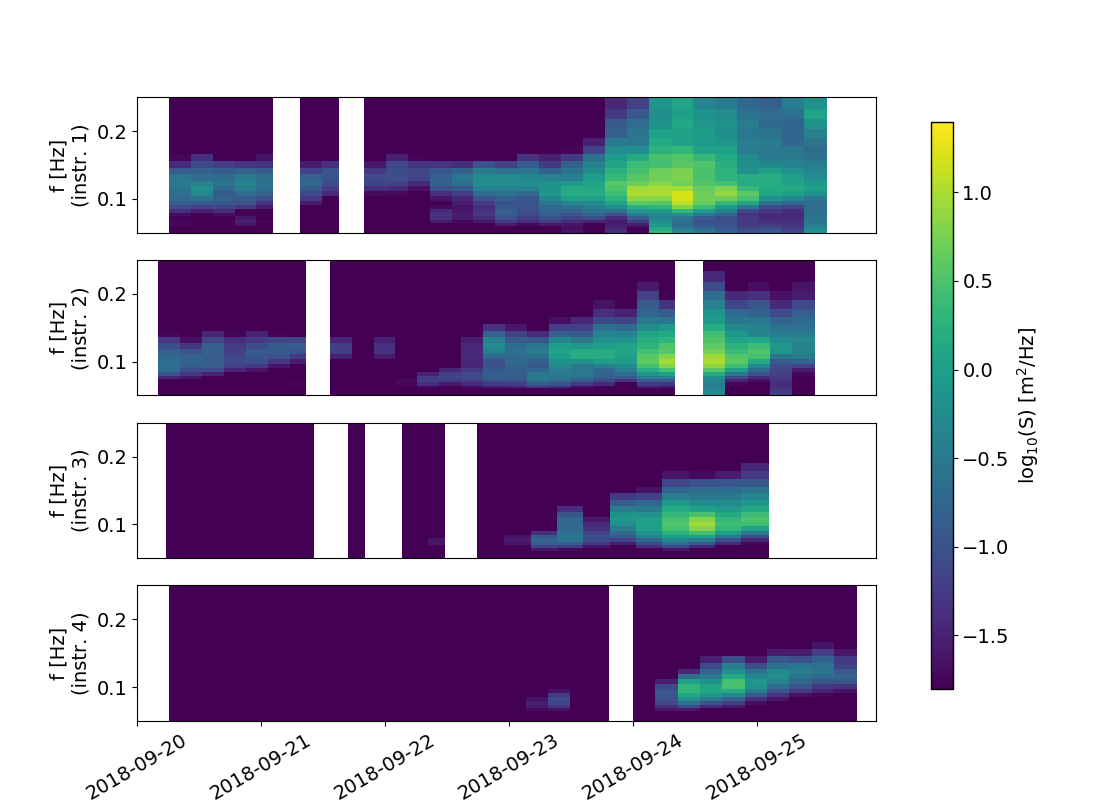}
    \caption{\label{fig_spectrograms} Spectrograms from all instruments, ordered going from the outer marginal zone (top) to the closed pack ice (bottom). Wave activity is clearly visible, and corresponds well with the duration of a storm in particular 24-09 to 25-09. Damping is clearly visible as the waves propagate deeper into the ice. This further validates the good functioning of the binning and transmission algorithm, and these data will be used for further calibration of waves in ice models.}
  \end{center}
\end{figure*}

The core mission of the instruments is to provide in situ measurements of waves in ice. The spectrograms presenting the wave information transmitted by the instruments are visible in Fig. \ref{fig_spectrograms}. Each spectrum is obtained using the Welch method, which is binned and compressed before being sent through Iridium, as described in the previous section. As visible in Fig. \ref{fig_spectrograms}, the patterns for the waves in ice activity are coherent among sensors. In particular, there is an episode of high wave intensity between September 24th and September 25th, which corresponds to a storm in the region. Clear damping is visible as the waves propagate deeper into the ice, and the damping is higher for high frequency waves in agreement to what is expected from theoretical considerations. These data further cross-validate the approach and algorithms used for the measurements of waves in ice.

\begin{figure}
  \begin{center}
    \includegraphics[width=.79\textwidth]{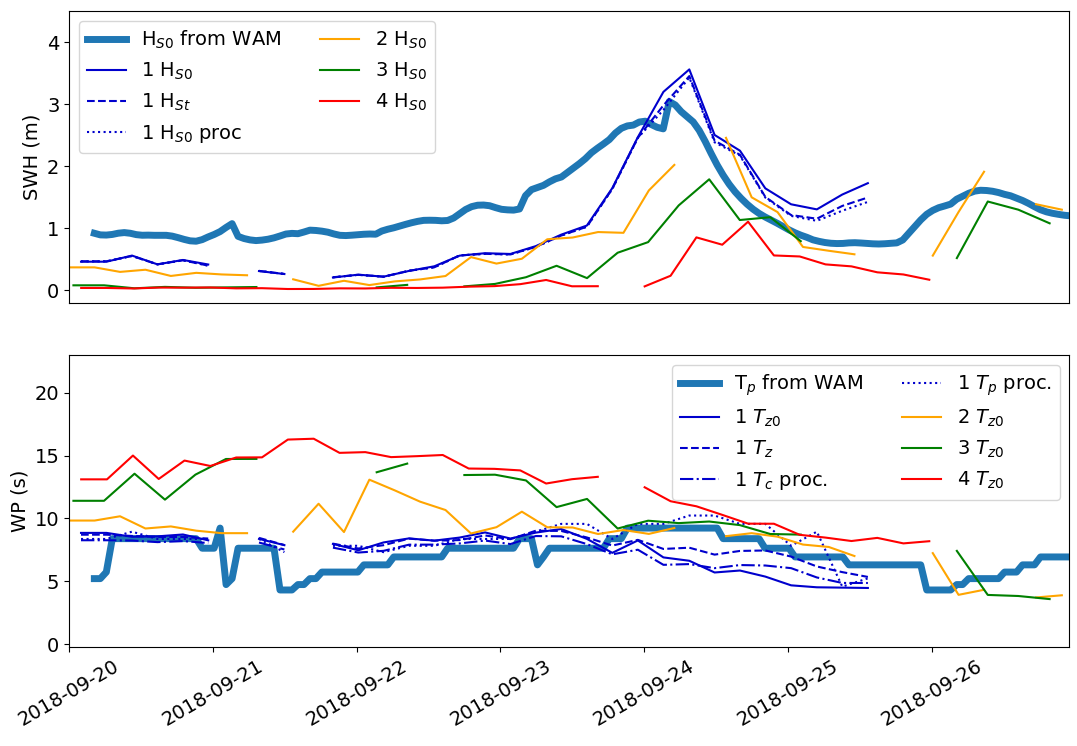}
      \caption{\label{fig:integrated_vs_transmitted} Comparison between the scalar quantities (significant wave height SWH and wave period WP according to different metrics, as described in section 2.2) transmitted by Iridium, the equivalent integral quantities calculated from the reduced spectra following the methodology presented in the previous section (quantities with suffix 'proc'), and results from the WAM waves model in the region close to the instrument outermost of the MIZ. The quality of the agreement between the different methods used for computing the wave height and period confirms both the methodology and under sampled spectra resolution chosen. In addition, the evolution of the wave properties as these travel through the MIZ are clearly visible. Namely, waves get attenuated (visible in the significant wave height), and the attenuation is larger for higher frequency waves (visible from the shift in the peak frequency of the waves). The quantities obtained from the reduced spectra are only presented for the first sensor to not overload the figure, but similar agreement is found for all instruments. Finally, reasonable agreement is obtained between the data reported by the instrument outermost in the MIZ and the predictions of the WAM model for the surrounding free-water wave field.}
  \end{center}
\end{figure}

Another cross-checking of the good quality of the on-board processing
and data compression can be performed by comparing the scalar results
transmitted, such as $H_{St}$, $H_{S0}$, $T_z$ and $T_{Z0}$, with similar
values calculated from the reduced spectra sent through Iridium. This
is illustrated for both quantities measuring wave height and period in
Fig. \ref{fig:integrated_vs_transmitted}. In this figure, the quantities
obtained from the scalar values transmitted through Iridium are presented
alongside those calculated from the reduced spectra (indicated by the 'proc'
suffix), following the methodology presented in the previous section. As
visible on Fig. \ref{fig:integrated_vs_transmitted}, we observe good
agreement between quantities transmitted and calculated from the reduced
spectra, which is an additional validation of both the methodology used and an
indication that the resolution of the reduced spectra is enough to capture the
dynamics of interest since the transmitted scalar quantities were computed
over the whole, non under-sampled spectra. This figure also further validates
the ability of the instrument to clearly detect changes in wave characteristics
arising as a consequence of their propagation through the MIZ. Namely, both
the wave damping and the preferential propagation of low frequency waves
are clearly visible from comparing the results obtained by the different
sensors. These data will be used to perform model calibration in later works.

Finally, the results obtained from the instrument further out of the
MIZ can be compared with wave model predictions in the region. For this,
we use the WAM wave model, run operationally by the Norwegian Meteorological
Institute. As developing models for wave propagation in the MIZ is challenging
and still a field of research, the model was set up so that ice with a
concentration over 3/10th is considered as continent and completely stops wave
propagation. Therefore, wave predictions are made only for instrument 1 and
ignore the effect of the sparse ice floes present in the region. Similarly,
the dynamic response of the ice floe is considered negligible to first
order, i.e. we assume that the floe and attached IMU mostly follow
the displacement of the water. The WAM wave model is a state-of-the art third
generation spectral wave model \citep{haselmann88}. The basic physics and
numerics of the WAM Cycle 4 wave model are described by \citet{komen94}. The
model solves the action balance equation, as do all third-generation spectral
wave models \citep{tolman91,booij99,ardhuin10}. In its current version, WAM
4.6.3 (freely available at \url{http://mywave.github.io/WAM/}), the source
function integration scheme of \citet{hersbach99} and the reformulated wave
model dissipation source function \citep{bidlot07} are used. The spatial
resolution is about 4~km, and the model covers the Norwegian Sea and the
Barents Sea on a rotated spherical grid with boundary conditions from the
European Centre for Medium-Range Weather Forecasts (see \citet{breivik09} for
details on the boundary scheme). The two-dimensional spectrum is represented
by 36 directions and 36 frequency bins, logarithmically spaced on 10\%
increments from 0.0345~Hz. The model is run daily to +66 hours. Here we
have concatenated forecasts from consecutive runs to provide a continuous
time series for comparison against our observations. The ice cover is taken
from the operational weather prediction model and is updated daily. There
will be discrepancies between the real ice cover and the modeled ice cover,
and the model has no way of distinguishing between fractional ice cover
and fast ice. For this reason, a hard ice boundary is set where the ice
concentration exceeds 30\%. Results are presented alongside the data from
the IMUs in Fig. \ref{fig:integrated_vs_transmitted}. As is visible there,
the agreement between the model and in situ measurements is generally
satisfactory considering the many coarse approximations used. This is one
more external validation of the good functioning of the instruments.

Regarding end of transmission, the battery level is clearly not to blame. Most
likely, environmental conditions were at the origin of both transmission
dropouts and finally loss of contact. The most probable causes 
are snow covering the antenna and shielding radio transmissions,
polar bears damaging the instruments (3 polar bears were encountered during
the deployment of the last two sensors), and ice breakup. This last factor is
especially likely for the sensor furthest out from the MIZ, which lost contact
when the storm was at its strongest. While little can be done about polar
bears and ice breakup, we will consider mounting the antennas on a pole on
the side of the main instrument in future iterations, in the hope that this
may help getting contact with the satellite. However, this constitutes a
tradeoff. Indeed, we believe that the reason why the instruments could survive
for such a long duration despite polar bear activity lies in the low profile
of the instruments, and probably a design featuring a case of larger vertical
dimension or a pole may be more attractive and get damaged withing short time.

\section{Conclusion and future work}

Instruments performing in situ measurements of waves in ice were successfully
designed in-house at the University of Oslo, tested and cross-tested
during short field campaigns, and deployed in the MIZ in the Arctic,
north of Svalbard during the Nansen legacy cruise in collaboration with
the Norwegian Meteorological Institute. High quality data on ice drift and
wave propagation in sea ice were obtained, which provides confidence in the
engineering solutions employed in designing the instrument. These data will
be used for further development and calibration of waves in ice models.

As both the hardware and software of the instrument are made available as an
open source design (see Appendix A), this opens new possibilities for in
situ measurements in the Arctic. Indeed, flexible and more affordable instruments may allow far more and higher quality measurements in harsh environments like the Arctic. Using open source code and PCBs allows for the reduction in the cost of the electronics to the point where the sensor performing the measurement represents over 50\% of the total cost. In addition, flexibility means that scientists will be able to quickly design the instruments they need by adapting the baseline design to their own needs, rather than going through long and costly contracting with private companies. This means that the production of a small series of instruments adapted to the needs of a specific measurements campaign can take place within a few weeks, therefore, greatly reducing risks and costs.

We will continue to work on this design to provide simplified assembly
processes for the end user and to further reduce costs and enhance
functionality in the years to come, adopting the same kind of incremental
refinements approach that was used for evolving from the first waves in ice
loggers presented in 2016 to the present instruments. In particular, we will work
further on isolating the IMU from magnetic disturbances induced by the battery and
electronics, so that also directional information can be obtained
in the future. We hope that sharing
our design may participate in creating an ecosystem for open source
instruments and benefit the community at large. Developing cost-effective
instruments for in situ measurements in the Arctic is a promising avenue for collecting the data that several communities need, and may help towards better studies of these challenging, remote environments.

\section*{Acknowledgement}

This study was funded by the Norwegian Research  Council  under  the
PETROMAKS2  scheme  (project  WOICE,  Grant  Number 233901, and project DOFI,
Grant number 28062). The Nansen Legacy (Arven etter Nansen) project
helped fund the wave loggers. We want to thank Kai H. Christensen and Matle M\"uller,
both from the Norwegian Meteorological Institute, for continuous support
and discussions and for offering the opportunity to join in the Nansen
Legacy Physical Processes cruise. Our warmest thanks go to the crew of
the RV Kronprins Haakon, for the time spent on the boat during the cruise.
{\O}B gratefully acknowledges funding by the Nansen LEGACY (Arven etter
Nansen) project, funded through the Research Council of Norway.
\section*{Appendix A: open source code and designs}

All the designs and files used for the building of the instruments,
including the PCB files ready for production, the code used on all
processors (low power unit, logger, Raspberry Pi), and general instructions
for assembling the instruments, are made available on the Github of the
author under a MIT license that allows full re-use and further development
(link:\url{https://github.com/jerabaul29/LoggerWavesInIce_InSituWithIridium}).
All software and designs are based entirely on open source tools, so that
the designs can be easily modified and built upon.

\bibliographystyle{igs}
\bibliography{bibliography}

\end{document}